%% file: template.tex
\pdfoutput=1

\documentclass[sensors,article,submit,pdftex,moreauthors]{Definitions/mdpi} 
\firstpage{1} 
\pubvolume{1}
\issuenum{1}
\articlenumber{0}
\pubyear{2024}
\copyrightyear{2024}
\datereceived{ } 
\daterevised{ } 
\dateaccepted{ } 
\datepublished{ } 
\hreflink{https://doi.org/} 

\Title{Galileo Project’s Observatory Class System Architecture}

\TitleCitation{Galileo Project’s Observatory Class System Architecture}

\Author{Phillip Bridgham $^{1}$, Alex Delacroix $^{1}$, Laura Domine $^{1,2}$, Andriy Fedorenko $^{1}$, Ezra Kelderman $^{1}$, Sarah Little $^{1}$, $^{3}$, Abraham Loeb $^{1,2}$, Robert Lundstrom $^{1}$, Eric Masson $^{1}$, Andrew Mead $^{1}$, Michael W Prior $^{1}$, Matthew Szenher $^{1}$, Foteini Vervelidou $^{1}$, Wesley Andrés Watters $^{1,4}$}

\AuthorNames{Phillip Bridgham, Firstname Lastname and Firstname Lastname}

\AuthorCitation{Bridgham et al.}

\address{%
$^{1}$ \quad Galileo Project, 60 Garden Street, Cambridge, MA, USA 02138\\
$^{2}$ \quad Harvard-Smithsonian Center for Astrophysics, 60 Garden Street, Cambridge, MA, USA 02138\\
$^{3}$ \quad Scientific Coalition for UAP Studies, Wellesley, MA 02138, USA\\
$^{4}$ \quad Whitin Observatory, Dept. of Physics \& Astronomy, Wellesley College, 106 Central St.
Wellesley, MA, USA 02481\\}

\abstract{Scientific investigation of Unidentified Anomalous Phenomena (UAP) is limited by poor data quality and incomplete data sets. Existing data are often fragmented, uncalibrated, and missing critical metadata. To address these limitations, the authors present the Observatory Class Integrated Computing Platform (OCICP), a system designed for the comprehensive scientific study of aerial phenomena which integrates multiple sensors to collect and analyze data on UAP. The OCICP system consists of two subsystems. The first is the Edge Computing Subsystem which directly interfaces with the sensors and is located within the observatory site. This subsystem performs real-time data acquisition, sensor optimization, and data provenance management. The second is the Post-Processing Subsystem which resides outside the observatory. This subsystem supports data analysis workflows, including commissioning, census operations, science operations, and system effectiveness monitoring. This design and implementation paper describes the system lifecycle, associated processes, design, implementation, and preliminary results of OCICP, emphasizing the system's ability to collect comprehensive, calibrated, and scientifically robust data. }


\begin{document}

\section{Introduction}
\input{sections/1-intro/introduction}

\section{Materials and Methods}
This section describes the high-level design and implementation features of the OCICP system to provide enough details to allow the open research community to replicate and build on the scientific approach presented to further the study of UAP within the realm of rigorous scientific inquiry.
\input{sections/2-material-methods/edge-design}

\input{sections/2-material-methods/edge-implementation}
\input{sections/2-material-methods/post-processing-design}
\input{sections/2-material-methods/post-processing-implementation}
\section{Results}
\input{sections/3-results/results}

\section{Discussion}

The results of the OCICP implementation demonstrate the system's capacity to collect calibrated, high-volume multimodal data suitable for scientific analysis of aerial phenomena. This system directly addresses the long-standing limitation in UAP studies: the lack of systematic, reproducible, and transparent data collection.

One of the most notable goals is the successful deployment of edge computing for real‑time object detection, tracking, and data fusion across sensor modalities. The Event‑Driven Architecture (EDA) and the implementation of the JDL Data Fusion Model have laid the groundwork for OCICP to identify, classify, and prioritize aerial events with minimal latency and high precision.

The post‑processing subsystem has proven essential in maintaining data provenance, paving the way for blinded data analysis, and supporting outlier detection through statistical and machine learning techniques. 

Despite these advances, several challenges remain. The complexity of the system, especially in coordinating data across diverse sensor types, presents integration and synchronization issues that require continual refinement. Furthermore, limitations in power and bandwidth at remote observatories constrain data transfer and require manual intervention.

Another challenge is the scalability of the post-processing pipeline, especially with the increasing volume of data generated daily. The success of ongoing and future deployments will depend on further optimization of HPC-based analysis workflows and streamlined data egress strategies.

Looking ahead, future work will focus on expanding sensor diversity (e.g., passive radar, optical spectrometers), enhancing classification algorithms through ensemble voting strategies, and further improving the dashboard interface for both research and support teams. We will deploy a front-end user interface for monitoring real-time tracking at the edge, enabling direct visualization of detections as they occur. In parallel, we will implement a comprehensive monitoring user interface for the entire OCICP system, which will allow researchers to review collected video footage, access aggregated analytics, and assess system status in general. Another major focus will be on enabling autonomous sensor response to aerial events and expanding real-time decision-making capabilities.

By contributing an open, FAIR-aligned, and reproducible system for the scientific study of UAP, OCICP serves as a model for the broader community. The adaptability, transparency, and scientific rigor of the platform offer a foundation on which other institutions and research efforts can be built.

\vspace{6pt} 




\authorcontributions{Conceptualization, P.B., A.D., L.D., A.F., A.L., E.M., M.P., and W.W.; Methodology, P.B., L.D., A.F., S.L., A.L., M.P., M.S, and W.W.; Software, P.B., A.D., L.D., A.F., E.M., and M.S.; Validation, A.D., L.D., A.F., S.L., E.M., M.S, and W.W.; Formal analysis, L.D., A.F., S.L., and W.W; Investigation, L.D., A.F., and W.W.;Resources, P.B., A.D., L.D., A.F., and E.M.; Data curation, A.D., L.D., A.F., R.L., A.M., F.V, and E.M.; Writing--original draft preparation, P.B., L.D., and A.F.; Writing--review and editing, P.B., A.D., L.D., A.F., E.K., S.L., A.L., R.L., E.M., M.P., M.S., and W.W.; Visualization, P.B., L.D., A.M., F.V, and A.F.; Supervision, L.D., A.F., A.L., and W.W.; Project administration, E.K, M.P., and A.L.; Funding acquisition, A.L.
}

\funding{This research was funded by private donations to the Galileo Project.}

\dataavailability{The raw data supporting the conclusions of this article will be made available by the authors on request.} 


\conflictsofinterest{The authors declare no conflicts of interest. The sponsors had no role in the design of the study; in the collection, analysis or interpretation of the data; in the writing of the manuscript; or in the decision to publish the results.} 


\begin{adjustwidth}{-\extralength}{0cm}

\reftitle{References}


\bibliography{refs}

%


\PublishersNote{}
\end{adjustwidth}
\end{document}

%% file: sections/1-intro/introduction.tex
\pdfoutput=1
The scientific study of Unidentified Anomalous Phenomena (UAP) is severely hindered by inadequate data collection and a lack of robust data. Current data, often collected by instruments not calibrated for scientific purposes, are fragmented, unsystematic, and lack sufficient metadata. Watters et al. discuss the motivations for investigating UAP through open, scientific study, including the limitations of previous research and the recent surge in public and governmental interest \cite{watters_scientific_2023}.

The Galileo Project is a multifaceted scientific research program active in the search for evidence of extraterrestrial technological civilizations (ETCs). This includes the search for artifacts, remnants, or potentially active objects in space as well as for scientifically anomalous objects near Earth. The ground-based field project described here aims to use custom designed telescope systems to detect and study unidentified aerial phenomena (UAPs) using scientific instruments in a variety of electromagnetic (EM) modalities such as optical, infrared and radio bands, in addition to audio and magnetic field signals \cite{loeb_overview_2022}. 

It should be noted that the main difference between the Galileo Project initiative and traditional SETI \cite{noauthor_seti_2024} approach to searching for extraterrestrial intelligence is that SETI focuses on the search for distant electromagnetic signals produced by ETCs, while this project concentrates on searching for scientific anomalies in or near Earth's atmosphere. We consider this approach to be complementary to the SETI approach.

\cite{watters_scientific_2023} provided a detailed description of the project's planned instrumentation, site selection criteria, and data processing strategies, highlighting the commitment to data transparency, data integrity, and the publication of findings in peer-reviewed journals. It laid out the overarching methodology and road map of which this system development effort is an integral part. 

This included a systematic scientific approach to studing anomalous aerial phenomena, using a multi-modal, multispectral system that encompasses various observational methods including optical and infrared imaging, radar, acoustic monitoring, and environmental data. The proposed methodology included a census of aerial phenomena, outlier detection, and hypothesis testing to determine whether there are aerial objects that represent phenomena currently unknown to science. The paper concludes with a detailed description of the planned instrument systems and their expected performance, the various sensor modalities required, data acquisition and archiving procedures, and the selection criteria for deployment sites. 

In this paper, we present the backbone of the system described above, our Observatory Class Integrated Computing Platform (OCICP).  OCICP is a pipeline for performing data reduction and producing analysis products that supports our system's primary scientific goals of determining if we can detect, identify, and characterize aerial phenomena  that are distinctly different from both currently known natural phenomena (e.g., birds, weather, meteors, etc.) and artifacts of human technological culture (e.g. aircraft, balloons, satellites, etc.). \cite{cloete_integrated_2023}. Our paper lays out the details of the design and implementation of system architecture, system life cycle, and associated processes for the OCICP element of the observatory class field station.

This paper uses Systems Modeling Language (SysML) views and diagrams \cite{noauthor_omg_2024} such as internal block diagrams for system context and activity diagrams to describe system behavior as conceptual tools to illustrate system elements and interactions. Although a SysML descriptive model has not been developed at this time, these system model-based engineering constructs serve to effectively communicate the system's structure and functions at a high level. Readers are encouraged to examine and refer to system diagrams when reading this paper, as the diagrams will provide the necessary context to better understand the goals, design, and implementation of the system.

This paper is organized as follows: In the Introduction section, we introduce background information relevant to understanding the need and application of the OCICP.  We show how the overarching system context and use case drive the subsystems and associated goals, as well as the system life cycle and primary processes.  Next, in the Materials and Methods section, we describe the design and implementation of both subsystems, including some key decision points for both design and implementation.  This content provides some details and insight to allow others to replicate and build on the OCICP architecture. In the Results section, we provide insight into the current status of the system, performance and effectiveness of this system architecture, technical challenges, results, and lessons learned.  The Discussion section explores several topics such as challenges with multi-sensor data fusion and scalability.  Finally, in the Conclusions section, we provide a brief summary of the overall paper.

\subsection{Motivation}
The lack of high-quality data is a recurring issue, as indicated by NASA's Independent Study Team Report. The independent study team report states that a new rigorous and evidence-based approach is needed that includes robust data acquisition methods and advanced analytical techniques \cite{nasa_nasa_2023, noauthor_dod_2023}.

On April 19, 2023, a congressional hearing was held that focused on the U.S. Department of Defense (DOD) efforts to understand and address UAP. At this hearing, the director of the All-domain Anomaly Resolution Office (AARO) reported to the Senate that they have 50 to 100-ish new reports each month and the primary reason for the unresolved cases is a lack of sufficient data, and without high-quality data, AARO cannot reach conclusions that meet rigorous scientific standards \cite{noauthor_dod_2023}, \cite{vergun_dod_2023}. 

Since the publication of Watters et al., there has been additional evidence of the difficulty and need to differentiate UAP from known objects or phenomena, such as the high volume of unresolved cases reported by the AARO, the government office responsible for investigating UAPs.

A challenge presented to providing data to the open research community is the fact that much of the data that is collected is classified because of the platforms and instruments that collected the data, not the data itself. For example, if an F-35 fighter jet took a picture of a bird, that data would be classified by default, not because there is anything unusual about the bird, but to avoid disclosing the capabilities of the F-35's sensors. To be clear, UAP phenomena are not classified, as NASA stated: "unidentified anomalous phenomena sightings themselves are not classified, it is often the sensor platform that is classified” \cite{donaldson_nasa_2023}. This constraint results in existing data often remaining inaccessible to the broader scientific community due to security concerns.

Collecting scientific data that are available for open research would facilitate collaboration between researchers, government agencies, and the public. This would bring a wider range of perspectives and expertise to the study of UAPs, potentially leading to a faster identification and understanding of these phenomena. This drive for open scientific data collection is also evident in the research efforts at JMU Würzburg highlighted by Kayal et al., which aims to establish a reliable database and foster interdisciplinary cooperation \cite{kayal_erforschung_2023}, and the UAPx collaboration described by Szydagis et al., which explicitly seeks to acquire high-quality, unclassified data and make it publicly available \cite{szydagis_initial_2024}. By addressing current limitations in data collection, introducing an open platform for collecting scientific data and embracing a culture of transparency; the scientific community can establish a more robust and evidence-based framework for investigating and understanding UAP  

\subsection{System Context and Use Case}
This initiative defines three classes of multi-instrument systems: they are Observatory systems, Portable systems, and Mesh systems \cite{watters_scientific_2023}. The Observatory system class is designed for the long-term study of UAP and is intended for comprehensive and continuous monitoring of aerial phenomena, aiming to create a detailed long-term census of objects in the sky within an Observatory site. The Portable systems class prioritizes portability, enabling deployment in different locations based on events or specific research needs. Portable Systems are less expensive than observatory-class systems and utilize a smaller suite of instruments, focusing on rapid deployment and data collection over shorter periods. The Mesh system class uses inexpensive and easily deployable sensors as part of a regional network to cover a wide geographical area \cite{watters_scientific_2023}. In the context of this paper, the term 'instrument' refers to a complete system or device that uses one or more sensors, often including signal conditioning, processing, display, and/or recording features.  The term 'sensor' refers to a device or component that detects a physical quantity (such as temperature, pressure, light, or motion) and converts it into a signal (often electrical) that can be interpreted or recorded.

Each of these system classes plays a crucial role in the multifaceted approach to UAP research, balancing cost, coverage, and data fidelity to effectively collect and analyze information about these enigmatic phenomena.  OCICP is an Observatory system class solution and the focus of this paper.  A diagram illustrating the system context of OCICP is shown in figure 1.

\begin{figure}[H]
\begin{adjustwidth}{-\extralength}{0cm}
\centering
\includegraphics[width=15.5cm]{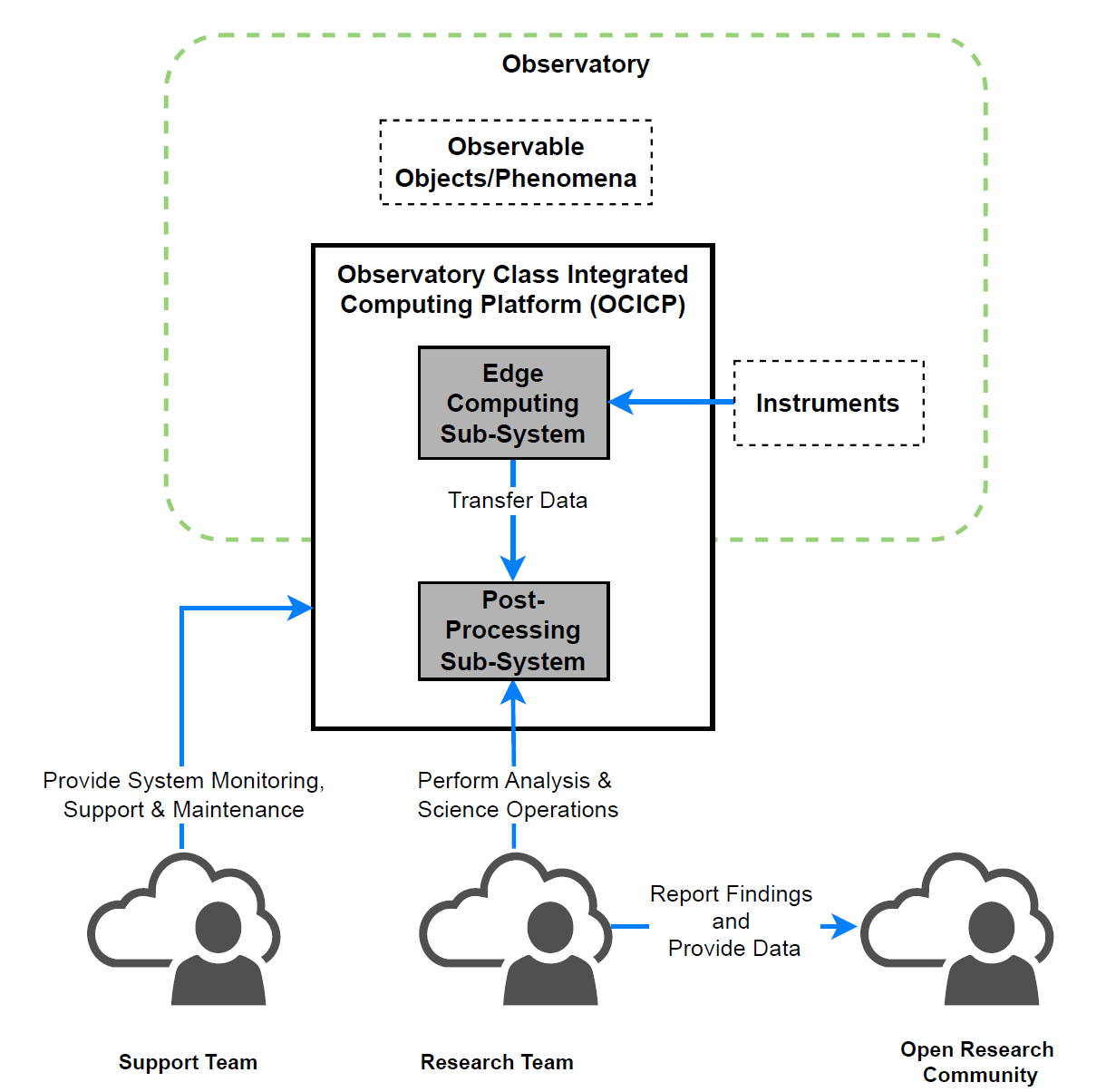}
\end{adjustwidth}
\caption{Observatory Class Integrated Computing Platform (OCICP) system context diagram depicting an observatory site, the OCICP two sub-systems, and the associated user communities.\label{fig1}}
\end{figure} 

The Observatory, depicted in figure 1 as a dashed green rectangle, is a ground-based installation of OCICP equipped with a suite of scientific instruments used to observe and collect data on UAP. OCICP is a system designed for the systematic and scientific study of UAPs that employs multiple sensors, artificial intelligence (AI), and rigorous data validation techniques. Observatories leverage instruments for multimodal coverage, including visible-light and infrared cameras, radio spectrum analyzers, microphones, and magnetometers \cite{watters_scientific_2023}. To support optimization of data collection, observatories use advanced data-processing techniques and AI to analyze the vast amounts of data collected by the instruments. This includes AI-based object detection and tracking in multiple modalities, instance matching, data fusion, and anomaly detection.

The primary user community for OCIPC is the Research Team.  The Research Team is composed of scientists with expertise in diverse scientific fields and is responsible for the project's strategy, technological choices, oversight of Commissioning, and execution of the Science Operations.  These processes are described in more detail in section 1.4. System Life-Cycle and Processes. The Open Research Community represents the broader scientific community outside of the research performed directly by this initiative. This initiative plans to publish its findings in peer-reviewed journals and make the data openly available to other researchers after following the Science Operations protocols described later in this paper. This open approach will allow independent corroboration of results and encourage further research in the field of UAP research. Refer to Watters et al. for more context on the role of the Research Team and the Open Research Community in the broader scientific study of UAP, where data transparency and independent corroboration of findings are emphasized \cite{watters_scientific_2023}.

The Support Team's role is centered upon keeping the Observatory up-time (actual operational time) as high as possible, which is accomplished by monitoring the operation of subsystems and performing routine and unplanned maintenance on all equipment and software.  In addition, the Support Team provides event-driven system performance assessment and tuning.

The OCICP is composed of two distinct subsystems, the Edge Computing subsystem and the Post-Processing subsystem.  These two subsystems, and their respective goals, are described next.

\subsection{Subsystems and Subsystems Goals}
The OCICP system must deal with a complex and diverse set of heterogeneous sensors and associated sensor modalities as described in the Science Traceability Matrix (STM), a foundational component of the Galileo Project UAP investigation \cite{watters_scientific_2023}. The breadth of sensing capacity is required to collect data adequate for the statistical analysis of rare outliers within the sea of conventional technological and natural phenomena. To accommodate the sensing, data collection, and requisite scientific analysis processes, the OCICP system has been designed around two subsystems: they are the Edge Computing subsystem and the Post-Processing subsystem. 

The Edge Computing subsystem is a set of computing devices designed and configured for potentially remote site deployment and are directly connected to the observatory's instruments. These computers control the instruments, as well as receive and record instrument data records. The goals of the Edge Computing subsystem are defined in the table below.

\begin{table}[H] 
\caption{Edge Computing Subsystem goals.\label{tab1}}
\begin{tabularx}{\textwidth}{>{\raggedright\arraybackslash}X>{\raggedright\arraybackslash}X}
\toprule
\textbf{Goal}	& \textbf{Description}\\
\midrule
Multi-Sensor Data Acquisition		& Support the collection of multi-spectral, multi-modal data from a configuration managed and calibrated sensor-based system.\\
\hline
Sensor and System Optimization		& Support the ability to optimize the use of the system and sensors for near real-time data collection opportunities.\\
\hline
Management of Data Provenance		& Support the immutability, integrity and traceability of sensor data and associated 
system and sensor configuration and calibration data for repeatable science operations.\\
\bottomrule
\end{tabularx}
\end{table}

With the goals related to data collection and data provenance allocated to the processing on the edge, the Post-Processing subsystem can be optimized for the support of science operations, which includes analysis of large datasets and utilization of high performance computing environments. \cite{watters_scientific_2023} describes the steps involved in the detection of outliers and hypothesis testing, and \cite{domine_commissioning_2024} shows the commissioning process of an all-sky infrared camera array that was enabled by the Post-Processing subsystem.

The Post-Processing subsystem is a hybrid computing environment for data science workflows and resides outside of the observatory. The goals of the Post-Processing subsystem are defined in the table below.

\begin{table}[H] 
\caption{Post-Processing Subsystem goals.\label{tab2}}
\begin{tabularx}{\textwidth}{>{\raggedright\arraybackslash}X>{\raggedright\arraybackslash}X}
\toprule
\textbf{Goal}	& \textbf{Description}\\
\midrule
Support the Commissioning and Census Operations		& While the commissioning process pertains to both the Edge Computing and the Post-Processing subsystems, the Post-Processing subsystem's main goal is to provide validation and processing of observatory data products as part of the census process. \\
\hline
Support the Science Operations		& The Post-Processing subsystem must support the processing environments, data management, and data science tools required by the Science Operations activities.\\
\hline
Support System Effectiveness Analysis		& The Post-Processing subsystem must support the ongoing system effectiveness monitoring activities, as well as support data analysis activities related to component calibration and evidence-based system design decisions. \\
\bottomrule
\end{tabularx}
\end{table}

The goals for the Post-Processing subsystem align directly to the primary operations supported by this subsystem, these are Commissioning and Census Operations, Science Operations, and Performance and Effectiveness Assessment. These operations are described in more detail later.

\subsection{System Life-Cycle and Processes}
The OCICP system life-cycle consists of four primary phases, they are: Development and Monitoring, Commissioning, Science Operations, and Decommissioning.  Across these phases are three types of activity: General, Site Specific, and Dataset Specific.  General activities refer to activities that are common across all observatories and observatory data sets.  Site Specific activities refer to activities that may be tailored and  must be conducted for each particular observatory and observatory data set. Finally, Dataset Specific activities are activities that are unique to a set of collected data.  The OCICP system life cycle is depicted in the following activity diagram.

\begin{figure}[H]
\begin{adjustwidth}{-\extralength}{0cm}
\centering
\includegraphics[width=15.5cm]{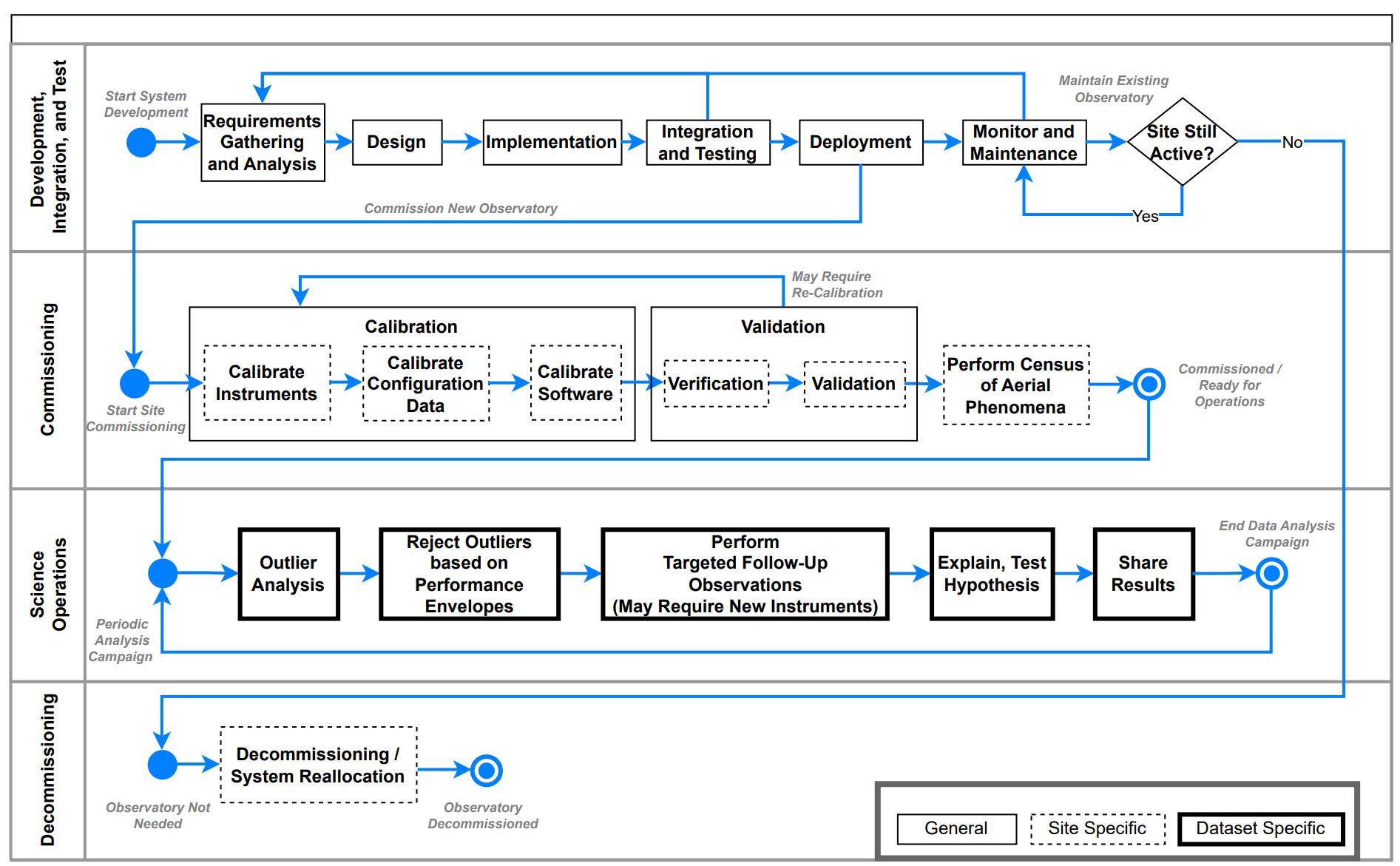}
\end{adjustwidth}
\caption{Observatory Class Integrated Computing Platform (OCICP) system life cycle diagram depicting the key activities across the four system life-cycle phases.\label{fig-lifecycle}}
\end{figure} 

The Development, Integration, and Test phase encompasses activities related to the system's development, deployment, and monitoring. Notably, this phase may lead to revisiting the Requirements Gathering step and then proceeding through the subsequent phases, enabling the management of latent system requirements and issue discovery. It is important to note that the Commissioning phase begins only after the Deployment phase concludes. Furthermore, the Development, Integration, and Test phases are cyclical and continue throughout the operational life of an observatory site.

The Commissioning phase of the OCICP system encapsulates the activities required for understanding the capabilities and limitations of sensors and instruments to support the refinement of the data analysis processes before executing the Science Operations phase.  The Commissioning phase may be executed for an entire OCICP system as well as for discrete sensor-based components within OCICP.

Refer to Watters et al. for project-level requirements for sensor calibration and system validation, which are key activities performed during the Commissioning phase \cite{watters_scientific_2023}.  Refer to Commissioning an All-Sky Infrared Camera Array for Detection of Airborne Objects \cite{domine_commissioning_2024} for a detailed discussion of our commissioning approach.

The Commissioning phase is primarily focused on preparing the system for full-scale scientific observation and operational readiness. During this phase, essential calibrations are conducted across sensors, configurations, and general system processes. The system's functionality is evaluated in the context of the overarching research objectives, ensuring alignment with the intended outcomes. In addition, this phase establishes the performance baseline and identifies system limitations, which are crucial for recognizing statistical outliers, formulating testable hypotheses, and applying likelihood-based statistical tests.

The Science Operations phase of the OCICP system refers to the continuous long-term data processing and analysis workflow as new data is collected. The Science Operations phase can only occur after the system has completed the Commissioning phase and has been verified to ensure the sensors and system have been built correctly, as well as validated, to ensure that the sensors and system produce reliable and scientifically useful results \cite{watters_scientific_2023}.  

The Decommissioning phase of the OCICP system involves the tear-down or reallocation of an observatory site.  This can happen for multiple reasons, including the termination of a land lease agreement or the desire to relocate an OCICP system.  When this occurs, the system must restart the system life cycle process at the beginning of the Commissioning Phase before Science Operations can begin.

Details of some of the activities within the OCICP system life cycle, including new methods and processes, are discussed in more detail in the next section.

%% file: sections/2-material-methods/edge-design.tex
\pdfoutput=1
\subsection{Edge Computing Subsystem Design}
\begin{figure}[H]
\begin{adjustwidth}{-\extralength}{0cm}
\centering
\includegraphics[width=15.5cm]{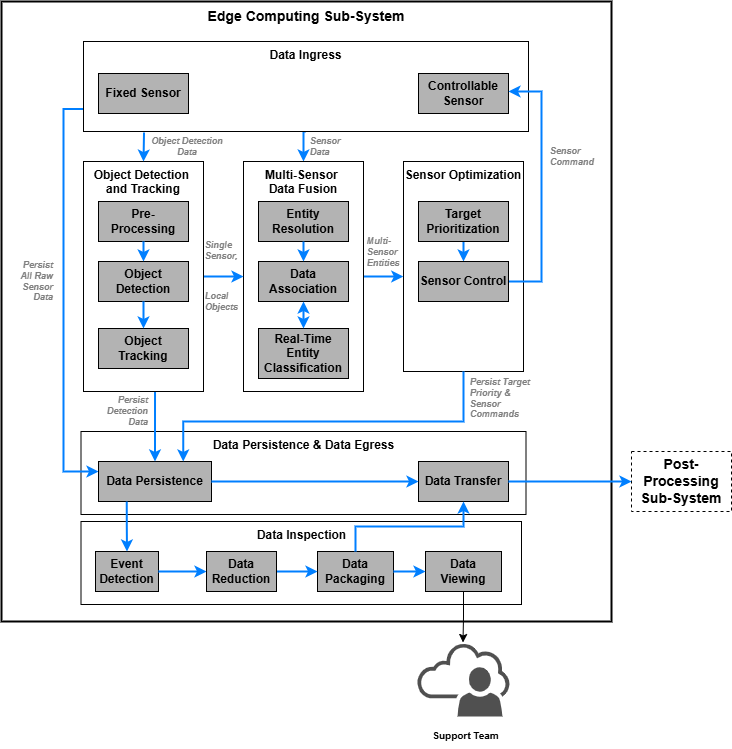}
\end{adjustwidth}
\caption{Edge Computing Subsystem design diagram depicting the functional decomposition of the Edge Computing Subsystem and the primary interactions. \label{fig-edge-design}}
\end{figure}

As mentioned above, the Edge Computing subsystem resides within the observatory site and has the goals of a) multi-sensor data acquisition, b) sensor and system optimization, and c) management of data provenance and integrity. \textit{Sensor and system optimization} represents the ability of the system to adjust data collection in real time based on observations.  An example of this is real-time targeting and tracking of an observed object.  The Edge Compute subsystem is composed of 5 functional areas: Data Ingress, Object Detection and Tracking, Multi-Sensor Data Fusion, Sensor Optimization, Data Persistence and Data Egress, and Data Inspection.  The design and functional decomposition of the Edge Compute subsystem is shown in Figure~\ref{fig-edge-design}. 

The container boxes represent the five functional areas of this subsystem, while the shaded boxes represent specific functions.  The arrows between the functions indicate the primary interactions between the functions, while the arrow heads indicate the primary source and target of the interaction.  Some labels are provided on the interactions to clarify the type of information that is shared for a particular interaction. 

Before describing each of the functional areas, this section will describe an architecture paradigm and a data processing framework that influences all functional areas.

\subsubsection{Architecture Paradigm: Event-Driven Architecture (EDA)}
The Edge Computing subsystem architecture has been designed from the ground up as an Event-Driven Architecture (EDA) system.  EDA is based on several core principles that differentiate it from other architectural styles \cite{luckham_power_2002}. In EDA, events represent the primary element. An \textit{event} is a significant change in the state of a system or an action taken by the system \cite{luckham_power_2002}.  The Edge Computing subsystem applies EDA design patterns for the implementation of event producers, event consumers, event channels, and event processing. The Edge Computing subsystem is designed to trigger responses based on events, making them key to interaction between services and components. In context of the Edge Computing subsystem, an event may represent a single instance a detection or observation, to more complex and correlated multi-sensor observations.

EDA has several advantages for a multi-sensor data collection system such as the Edge Computing subsystem.  EDA enables asynchronous processing, which allows better handling of data spikes and large volumes of requests. The application of EDA design patterns, namely event producers, event consumers, and event channels, results in a loose coupling between components.  This design approach improves the extensibility and flexibility of the system, which is critical to supporting the integration of heterogeneous sensors and data processing techniques in the system. Finally, as a near real-time event-based system, EDA enables the Edge Computing subsystem to respond immediately to changes, enabling real-time decision-making. All system clocks are synchronized with GPS time, as well as monitored for any significant drift. The key attributes of the Edge Compute subsystem are described next, organized by functional area as presented in Figure~\ref{fig-edge-design} from top to bottom.

\subsubsection{Functional Area: Data Ingress}
The Data Ingress functional area consists of all sensors.  Within this functional area, two types of sensors exist, sensors that are controllable and sensors that are not.  At this time, there are minimal external data sources utilized by the Edge Compute subsystem, primarily for systems operations purposes such as sunrise/sunset schedules.  The closest data type the OCICP has to an external data source is the ADS-B data; however, these data are collected via a local software defined radio tuned for ADS-B data collection and therefore are considered a local sensor.  All data, for both types of sensors, are persisted by the Data Persistence and Data Egress functions.  This is indicated on the left side of the design diagram.  Controllable sensors can be controlled by the Sensor Optimization functions, as indicated on the right side of the design diagram. By ``controllable sensor" we are referring to sensors that can be directed to respond to events during observation, as compared to less sophisticated sensors restricted to predefined behavior. Refer to Watters et al. \cite{watters_scientific_2023} for more details on the role of different sensor types in the overall data collection strategy.

\subsubsection{Functional Area: Object Detection and Tracking}
The Object Detection and Tracking functions perform object detection and tracking for sensor data capable of providing object identification and/or position information. Although traditional computer vision definition may be limited to the processing, analysis, and interpretation of visual data from images and video sensors, our definition expands on this.  Computer Vision here may include audio, spectral, radar, and other sensor types that may aid in capturing information about the world and aids in extracting meaningful information for interpreting objects. 

The three steps of Pre-Processing, Object Detection, and Object Tracking represent a generalization of computer vision steps. Depending on the sensor modality and features, observation events with object identification and/or position data will be routed through as many of the three steps as the information is capable of supporting. Pre-Processing represents activities such as data acquisition, noise reduction, and other activities that prepare the data for analysis.  Object Detection starts with distinguishing objects from the background scene, including establishing an initial position of the object from the scene. This can be accomplished with a single detection or with a temporal sequence of detections. Finally, Object Tracking is the process of identifying and continuously monitoring the location and trajectory of a specific object (or multiple objects) in a sequence of video frames.  

Events produced from the Object Detection and Tracking function represent local, or single sensor, objects with features and optional location data. In addition, this function may return a set of consecutive object positions or a track with an associated local object identifier. The Edge Processor's ability to perform real-time object tracking is separate from object track analysis performed within the Post-Processing subsystem. 

\subsubsection{Data Processing Framework: Data Fusion Model}
Data fusion is the process of combining data from various sources to refine state estimates and predictions. This involves determining relationships among the data (data association) and using the combined information to improve the estimations of an environment or situation (state estimation). Heterogeneous sensor modalities may have their own distinct representation, for example, optical data are a times series of 2-D color images, acoustic data are a 1-D time series of voltages representative of sound pressure level, and radio data are a times series of amplitudes of frequency-binned power spectra. The Data Fusion and Exploitation (DFE) system operates at a level of abstraction that treats all incoming data generically, extracting fundamental parameters such as detections, tracks, identities, and other key features from each individual data point.  Data fusion aims to refine and predict the states of entities, groups of entities, and their relationship to system objectives. The concept of data fusion can be applied to various domains, ranging from military applications such as target tracking and awareness of battlefield situations to civilian applications such as medical diagnosis, weather forecasting, and more \cite{steinberg_revisions_1999}.

For OCICP, applying the Data Fusion Model conceptual framework allows for the combination of data from these different sources to provide a more comprehensive and accurate understanding of observed phenomena, as opposed to relying on unconnected, individual sensors each with limited perspectives.  The OCICP system incorporates the Joint Directors of Laboratories (JDL) Data Fusion Model \cite{steinberg_revisions_1999}, as well as expansions to the model \cite{ii_handbook_2017}.  This model is a widely used method for categorizing functions related to data fusion. The following table describes the five levels of the Data Fusion Model with examples of how these levels align to functions that benefit the goals of the Edge Computing Subsystem. This paper uses Data Fusion Model definitions for the terms "object" and "entity". The Data Fusion Model defines an "object" as a reference to physical or observable phenomena in the context of sensor data, while the term "entity" is a broader concept often implying grouped or contextualized objects. For the OCICP system, objects are representations in each modality, and entities are real-world objects.  The ability to translate the observable objects into entities is a key goal of a data fusion solution.

\begin{table}[H] 
\caption{JDL Data Fusion Model and OCICP Scenario Alignment.\label{tab3}}
\begin{tabularx}{\textwidth}{>{\raggedright\arraybackslash}X>{\raggedright\arraybackslash}X>{\raggedright\arraybackslash}X} 
\toprule
\textbf{Data Fusion Level} & \textbf{Data Fusion Model Functionality Definition} & \textbf{OCICP Example (Not Necessarily Implemented)} \\
\midrule
Level 0: Sub-Object Data Assessment (Signal/Pixel-Level Fusion) & This level focuses on processing raw sensor data at a pixel or signal level to reduce noise and extract basic features. & Data from the object detection cameras consist of images and metadata for objects detected, bounding box coordinates within image pixel matrix and local ID for tracking. \\
\hline
Level 1: Object Assessment (Object-Level Fusion) & This level detects, identifies, and tracks individual objects or entities within the data. This level fuses information from different sensors. & Objects detected by detection cameras are correlated with ADS-B receiver data using camera extrinsic calibration to compute physical 3D position and determine if the detected object is an airplane. \\
\hline
Level 2: Situation Assessment & This level provides an understanding of the relationships between objects and the environment to determine what is happening in the scene. & Audio from audio sensors indicates an airplane is present. This data applies to the area of interest and contextualizes events within this same event window. \\
\hline
Level 3: Impact Assessment & This level predicts the potential outcomes or consequences of the current situation. & Entities that have been classified as an airplane have their target prioritization adjusted accordingly (e.g., priority lowered). \\
\hline
Level 4: Process Refinement (Resource Management) & This level ensures that sensor systems and computational resources are used effectively. & The PTZ (Pan, Tilt, Zoom) camera is sent new commands that correspond to the latest prioritized targeting opportunities. \\
\hline
Level 5: User Refinement (Cognitive Refinement) & This level focuses on the interaction between the data fusion system and human operators, integrating user feedback and cognitive inputs to refine the system’s performance. & Data is recorded and available for review for interpretation of events and analysis of system performance and effectiveness. \\
\bottomrule
\end{tabularx}
\end{table}

The scenario described in the table above illustrates how the JDL Data Fusion Model aids in aligning system functions to enable the Edge Computing Subsystem to meet the goal of Sensor and System Optimization. Though there exists a single Multi-Sensor Data Fusion functional area within the center of the system design diagram, we can see that the functional levels of the Data Fusion Model are applied across multiple Edge Compute subsystem functional areas.  As an example, Level 0 is contained within the Object Detection and Tracking functional area, while Levels 3 and 4 are contained within the Sensor Optimization functional area. 

\subsubsection{Functional Area: Multi-Sensor Data Fusion}
The Multi-Sensor Data Fusion functional area receives single sensor observation events and performs Level 1 of the Data Fusion Model, Object Assessment. The first step within this functional area is to perform entity resolution. Entity resolution is the process of identifying and merging features of objects that refer to the same real-world entity.  On first occurrence of this action, a single-sensor object is promoted to a multi-sensor entity.  Entity resolution is primarily executed within Level 1 of the Data Fusion Model.  The second step is to complete data association or data fusion across all available sensors as new events are processed and objects are resolved as potential multi-sensor entities.  Finally, the third and last step of Real-Time Entity Classification attempts to classify entities based on all state estimations for an entity.  The output of the Multi-Sensor Data Fusion function are events representing potentially multi-sensor entities, the current location of the entity, and an estimation of the entity’s class (classification).

\subsubsection{Functional Area: Sensor Optimization}
The Sensor Optimization functional area performs Levels 2 and 3 of the Data Fusion Model.  At this stage, sensor data have been correlated across individual sensors and associated with entities, where possible, are available for environment/scene contextualization.  These actions are part of Level 2: Situation Assessment.  The Target Prioritization function utilizes this real-world state estimation to prioritize real-time observable objects as part of the Level 3 Impact Assessment.  Part of the Edge Computing subsystem goals are optimizing sensors to capture high-priority objects. The output of the Target Periodization function is a targeting event.  The targeting event is then processed by the Sensor Control function, which is responsible for translating general targeting needs into specific controllable sensor commands.

The Data Inspection functional area provides capabilities for the packaging, viewing, and evaluation of data, including system performance evaluation related to detected events.  This functionality provides the opportunity for both process and cognitive refinement as defined by the JDL Data Fusion Model.

\subsubsection{Functional Area: Data Persistence and Data Egress}
The Data Persistence and Data Egress functional area is responsible for saving sensor and sensor meta data, as well as data providing transfer support to transfer data from the observatory site to the Post-Processing subsystem data lake. For the Edge Computing subsystem, Data Persistence is provided for all raw sensor data, as well as logging of events such as object detections and sensor control commands.  These activities are indicated by the arrows that go into Data Ingress from Data Persistence, Object Detection and Tracking, and Sensor Optimization.

\subsubsection{Functional Area: Data Inspection}
The Data Inspect functional area satisfies Level 4 and Level 5 of the Data Fusion Model.  For Level 4, optimizing resource management, analysis of detections by the Event Detection function can result in performance and effectiveness measurements, and in turn be used to assess and refine the system.  For Level 5, cognitive refinement, the Data Packaging and Data Viewing functions together provide the ability to review the data support interactions with human operators and aids in situational understanding of events.

%% file: sections/2-material-methods/edge-implementation.tex
\pdfoutput=1
\subsection{Edge Computing Subsystem Implementation}
This section provides a more detailed view of the Edge Computing subsystem, with a focus on its implementation and component interactions to illustrate the system's dynamic behavior. Some implementation features described in this section may not be in use currently, or even operational, due to the where we are in the prototyping, development, and commissioning cycle for each instrument; however, all of the implementation details described in this section represent capabilities that have been developed and have been used at some time, unless otherwise specified as 'planned'. 

\begin{figure}[H]
\begin{adjustwidth}{-\extralength}{0cm}
\centering
\includegraphics[width=15.5cm]{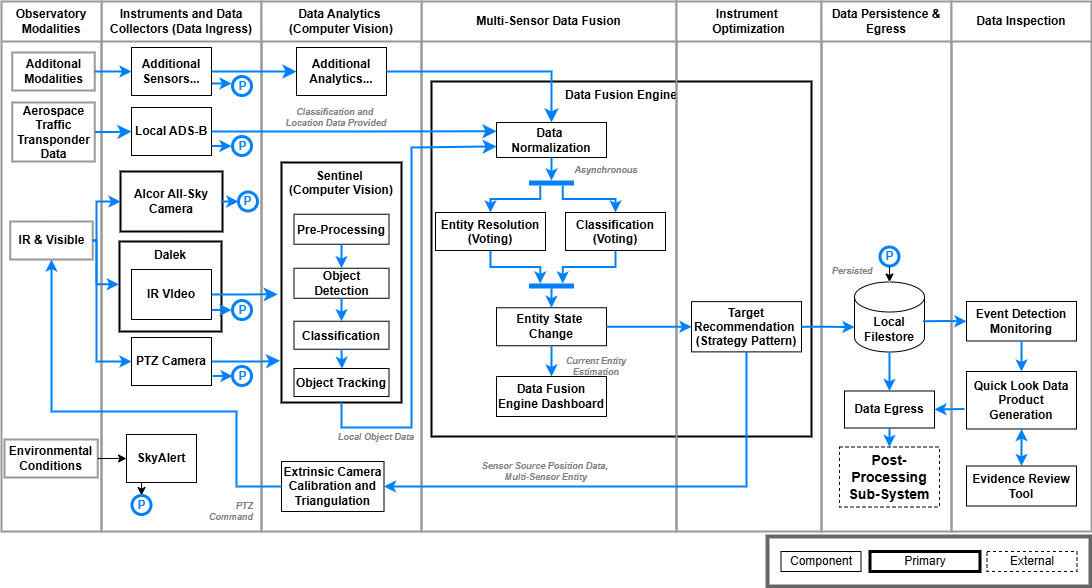}
\end{adjustwidth}
\caption{Edge Computing Subsystem Component interaction diagram depicting the implemented components and primary interactions, organized by capability categories.\label{fig5}}
\end{figure} 

The interaction diagram above in Figure~\ref{fig5} depicts the elements of the Edge Computing subsystem as primary components (bold boxes), components (boxes), and component interactions (blue lines), all of which are organized within the component categories (columns).  Elements external to the Edge Computing subsystem are indicated with a dashed border. For the most part, the component categories in this diagram match directly to the functional areas described above; however, several of the component categories here have alternative names to better represent their function. As described above, the Edge Computing subsystem is designed around the EDA architecture paradigm.  The messaging in the Edge Computing subsystem is implemented using the ZeroMQ messaging library \cite{zeromq_zeromq_2023}.  ZeroMQ is a high-performance asynchronous messaging library that provides a simple API for building scalable, distributed, and concurrent applications by abstracting away low-level socket management and enabling various communication patterns such as publish/subscribe, request/reply, and push/pull.  Using OpenJDK Java \cite{noauthor_openjdk_2024}, the Data Collector design pattern was established to provide a reusable Java class template to provide common data collection services.  Unless specified otherwise, all data collectors are implemented using this Data Collector class, and all component interactions are implemented using ZeroMQ.

\subsubsection{Observatory Modalities}
The Observatory category represents the area of interest for which the Edge Computing subsystem’s sensors are monitoring. Components within this category include observable objects and observable phenomena, including environmental conditions. The Additional Phenomena block represents future observable phenomena requirements that may be added to the Edge Computing subsystem sensing capacity.  An example of this includes optical spectroscopy and passive radar object localization as described by Randall et al. \cite{randall_skywatch_2022}.
\subsubsection{Instruments and Data Collectors}
The Instruments and Data Collectors category represents the sensing and data collection capacity of the Edge Computing subsystem.  At the time of this writing, the Edge Computing subsystem has the ability to sense local aerospace traffic, detect objects using IR cameras, detect objects emitting sound, capture video with a PTZ (Pan, Tilt, Zoom) optical camera, and collect environmental data.   
Local ADS-B (Automatic Dependent Surveillance–Broadcast) \cite{faa_automatic_2023} data is captured using a Raspiberry Pi Model 4 \cite{raspberry_pi_raspberry_2024} with a FlightAware ProStick® Plus \cite{noauthor_flightaware_2024}.  A Data Collector was created to receive data from the FlightAware software, maintain the state of all observable airplanes, and publish ADS-B notification for the Data Fusion Engine to process. 
This category contains the ability to detect aerial objects from optical, infrared (IR) cameras utilizing the custom Dalek system and the Alcor All-Sky camera, as well as the ability to detect aerial objects that emit sound.  The hardware and software design of this part of the system is described in previous work \cite{szenher_hardware_2022}. 
The collection of environmental data is implemented using SkyAlert \cite{noauthor_skyalert_nodate}.  For the Edge Computing subsystem, environmental data such as ambient temperature, humidity, wind speed, barometric pressure, cloudiness, etc. can provide contextual information related to observable phenomena, and may influence the Level 2, situation assessment, as well as other levels of data fusion.  Additionally, for the Post-Processing subsystem, environmental data will be used as part of commissioning and the outlier detection analysis. 
All data from sensor data collectors persist, as indicated with the circled P icon.  The Additional Sensors block represents future sensor and data collection capabilities that may be added to the Edge Computing subsystem. An example of this is the implementation of sensing and data collection for a magnetometer.

\subsubsection{Data Analytics}
The Data Analytics category is a collection of capabilities that perform post-processing, enrich, and / or prepare sensor data for data fusion. A significant component of this category is the custom Sentinel software developed by the Galileo Project for camera-based object detection, object localization, and object tracking which leverages the ML based real-time object detection software frameworks. \cite{jocher_yolov5_2020}. Sentinel processes video frames provided by a Gstreamer \cite{noauthor_gstreamer_2024} pipeline. At this time, Sentinel is written in Python v3 \cite{python_software_foundation_python_2008}, however, efforts are underway to translate Sentinel to C++ to improve real-time performance. The initial design of Sentinel was described in \cite{cloete_integrated_2023}. Details of the design and implementation of Sentinel for inference of the kinematics of objects in the sky based on their tracks are discussed in \cite{szenher_hardware_2022}.

The Extrinsic Camera Calibration and Triangulation functionality is shown at the bottom of the Data Analytics category.  This functionality represents the ability to triangulate the 3D position of an object using multiple camera views.  This technology is written in Python and can be used within the event flow between Target Recommendation and the PTZ camera within the Edge Computing subsystem.  Additionally, this Python library can be used within the Post-Processing subsystem. The approach and implementation details for this are described in depth in \cite{szenher_hardware_2022}.

\subsubsection{Multi-Sensor Data Fusion}
The Multi-Sensor Data Fusion category is realized by the DFE (Data Fusion Engine), which is based on a custom Java Spring Boot framework \cite{pivotal_software_spring_2014} web application with multi-threaded message processing capability. As an event-based solution, the DFE applies the Publish-Subscribe design pattern \cite{hohpe_enterprise_2003} and can be configured to subscribe to various message topics.  For performance and scalability reasons, all message subscribers are implemented as a multi-threaded process using a configurable thread-pool.  Message processors are implemented using factory and strategy design patterns \cite{gamma_design_1994} to maximize extensibility, while the messages themselves are implemented using the Data Transfer Object (DTO) \cite{alur_core_2003} enterprise design pattern.

The DFE can be configured for multiple object detection topics. As an example, the Dalek instrument provides eight camera feeds, each of these would be specified in the DFE configuration file, including the messaging host, topic, and port \cite{zeromq_zeromq_2023}.  Available video stream URLs may also be specified for display via the DFE Dashboard (discussed later).

Within the context of data fusion, a local object is a sensor-specific observation of an entity within the sensor's immediate environment, while a global entity is a fused, system-wide representation of that entity derived from local observations from multiple sensors. The DFE maintains a list of both local objects and global entities by continuously collecting, timestamping, and associating local observations with their corresponding global entities within a defined time window for dynamic updating and correlation. To this end, the DFE maintains the following entity maps:
\begin{itemize}
\item	a map of global entities, with key look-up based on global ID
\item	a map of local objects, with a key look-up based on instrument/sensor ID and local ID
\item	a map of global entities to local objects, with a key look-up based on global ID, supports one-to-many associations
\item	a map of classification votes, with individual voting records
\item	a map of ADS-B records, as received from the ADS-B local sensor source
\end{itemize}

In total, this state management represents the best state estimate from DFE in the real world view.  As these states change, the DFE publishes events of entity state change across all sensor sources. To manage the state management in-memory footprint, a configurable purging policy is in place. The purging policy can be configured in terms of frequency of the purge check and entity retention time after the last observation event.  When the criteria are met, the purge policy executor will enforce the policy and clean up records across all maps in a thread-safe process.  With this policy strategy, the DFE can be configured to 'forget' about an entity after a configurable amount of time has passed since it was last observed.

At the core of the DFE is Entity Resolution, which was defined earlier within the topic of Data Fusion Model.  It is anticipated that the design strategy for entity resolution will evolve over time and perhaps even include multiple strategies depending on sensor modalities.  To support this scenario, the DFE applies the Strategy Pattern \cite{gamma_design_1994} to determine the most effective implementation of entity resolution based on the type of observation message. The DFE allows for the configuration of multiple strategies and the association of these strategies for event source types (or sensor types). This approach enables the DFE to determine, in real-time, the optimal approach to use for entity resolution based on the sensor modality and other considerations. To support multiple current strategies, or ensemble approaches to entity resolution, a voting mechanism is implemented.  The voting mechanism will allow multiple techniques to be applied.  The currently used strategy is based on a temporal and location comparison with configurable tolerances.  This allows for the resolution of entities across multiple sensors where latitude, longitude, altitude, and time are provided.  It is anticipated that different sensors and sensor hosts may have varying degrees of accuracy, and so the DFE may be configured for confidence weights in addition to tolerances per sensor type.

A long-term goal of the DFE is to provide real-time classification of objects to support the Target Recommendation step within the Sensor Optimization category.  This feature is not implemented currently; however, the design calls for the application of a voting mechanism, as in the case of Entity Resolution. The voting mechanism will allow votes to consist of one or more chosen classifications, a confidence score, and a source/calculator weight.

The Data Fusion Engine Dashboard is discussed later within the Data Inspection category.

\subsubsection{Sensor Optimization}
The Sensor Optimization category is implemented as the target recommendation logic that can be leveraged within the DFE via the configurable Strategy Pattern.  Sensor optimization is fulfilled by DFE because it is desirable to keep the event processing across data fusion and sensor optimization as efficient as possible.  By implementing target recommendation logic as Java classes that can be configured to be executed within DFE, inter-component communications were avoided. In other words, if the target recommendation were implemented as an additional and external application, then the DFE and the target recommendation applications would need to share a considerable amount of time-sensitive data. By embedding the target recommendation feature into the DFE, the message overhead was avoided. The DFE supports the configurable selection of target recommendation strategies. Our initial implementation of target recommendation logic will send pointing commands to the PTZ camera, as well as optical zoom settings.  In the future, we envision several other applications such as the pointing of directional microphones, configuration and tuning of spectrometers, etc.

\subsubsection{Data Persistence and Egress}
The Data Persistence and Egress category is a logical representation of distributed file writing by various components.  In most cases, data are written to a file at the source of the data generation/collection.  In some cases, this component responsible for data generation/collection may have limited resources or significant performance requirements, and may utilize a threaded ZeroMQ publisher library to 'fire and forget' data.  In this situation a Data Collector will be running to subscribe to these messages and perform the data persistence, perhaps on a different host with more resources.  The points at which data persistence is required are noted as a circled P within in figure 6.  Currently, all data are persisted as flat files, and no databases are used on the Edge Computing subsystem.

The file storage strategy incorporates mechanisms to retain information on the source and traceability of data.  The figure below illustrates the file name and file path conventions used to capture important information about the data within the files.
\begin{figure}[H]
\begin{adjustwidth}{-\extralength}{0cm}
\centering
\includegraphics[width=15.5cm]{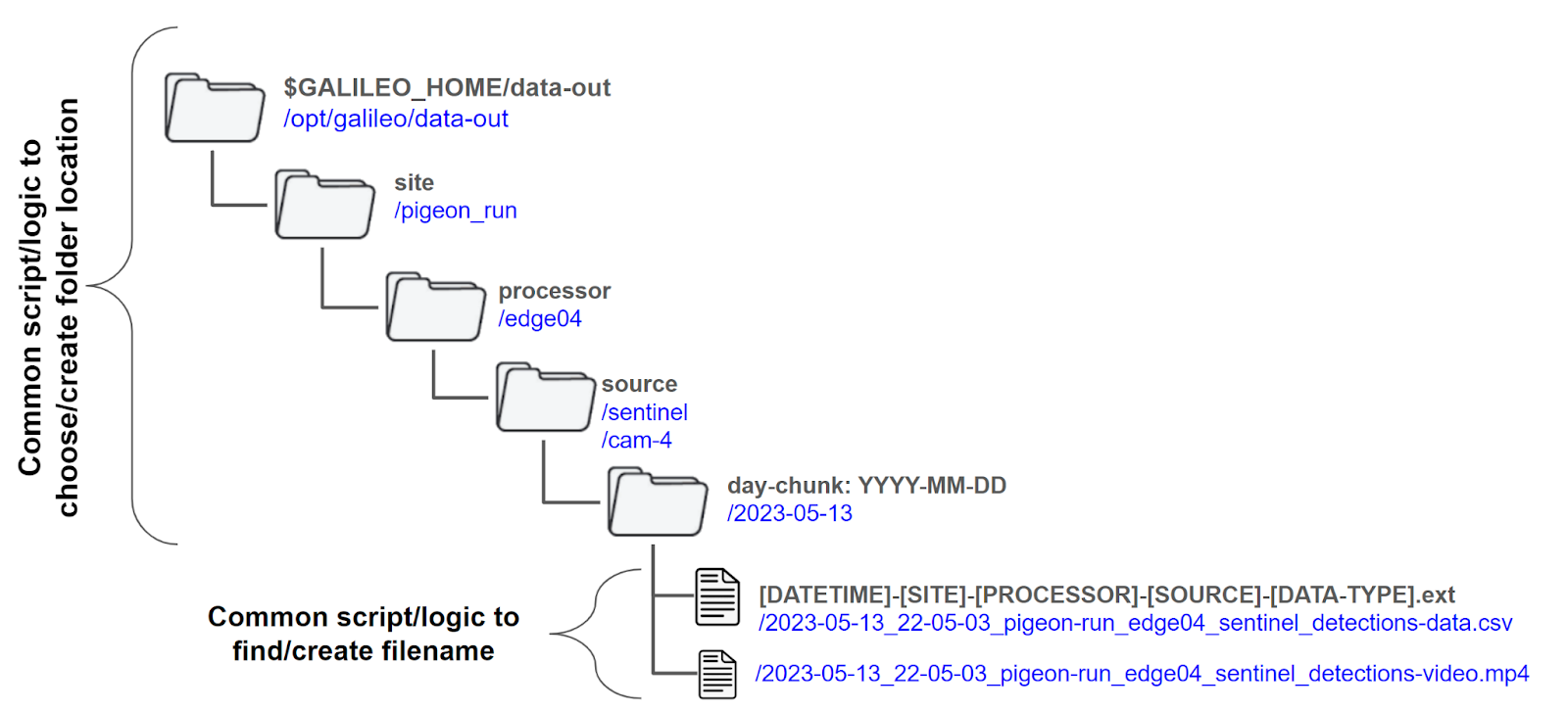}
\end{adjustwidth}
\caption File storage path and naming convention for structured data traceability. The hierarchy captures key metadata including site, processor, source, and date, while standardized file names encode essential attributes for automated processing and retrieval.\label{fig7}
\end{figure}
The file storage strategy utilizes an environment variable to establish the root folder location, from here several conventions are applied to establish final file location for storage and retrieval, they are:
\begin{itemize}
\item Output folder: all data production is placed under here, simplifies data egress processing by providing a single parent folder for data extraction
\item Site: organizes data by observatory moniker, helps manage the collection of  multi-observatory data for the post-processing environment
\item Processor: the host name for which the data was generated/collected
\item Source: the id of the instrument or sensor of the source of the  data file
\item Day-chunk: a folder to separate data files by date of generation, supports incremental data egress processes and efficient data look-up by date range
\end{itemize}

In addition to the folder conventions, a data file naming convention has been established to encode the above data, as well as a data-type value that represents the type of data that is contained within the file.  This file naming convention enables traceability of data, as well as manual and programmatic file filtering based on date, site, host, source, and data type.  A reusable Python and Java class was created to help leverage and enforce these conventions. 

Data Egress is implemented via a combination of custom software and scripts.  Most data files are generated in a common system-configured location with a structured folder hierarchy according to source/host, sensor type, and date/time.  Data formats of interest are discussed in the Data Inspection section below.

\subsubsection{Data Inspection}
The need and opportunities for data inspection have been discussed earlier in the design section.   To this end, the DFE provides a web-based dashboard that supports the following features:
\begin{itemize}
\item	Role-based use authentication and authorization
\item	Pre-configured dashboards (b)
\item	Configuration-driven layout and real-time graphical display of detection cameras with 2D representation of current detection status (c)
\item	Tabular display of detection camera records with global IDs(e)
\item	Tabular display of real-time ADS-B records, with global IDs and aircraft information, decorated with current target selection (d)
\item	A real-time geospatial display of entities with real-world coordinates (currently limited to ADS-B entities) (f)
\item	User authentication and authorization for control of access

\end{itemize}
These features are annotated in the DFE Dashboard screen capture in Figure~\ref{fig7}, except for A (the login screen).

\begin{figure}[H]
\begin{adjustwidth}{-\extralength}{0cm}
\centering
\includegraphics[width=15.5cm]{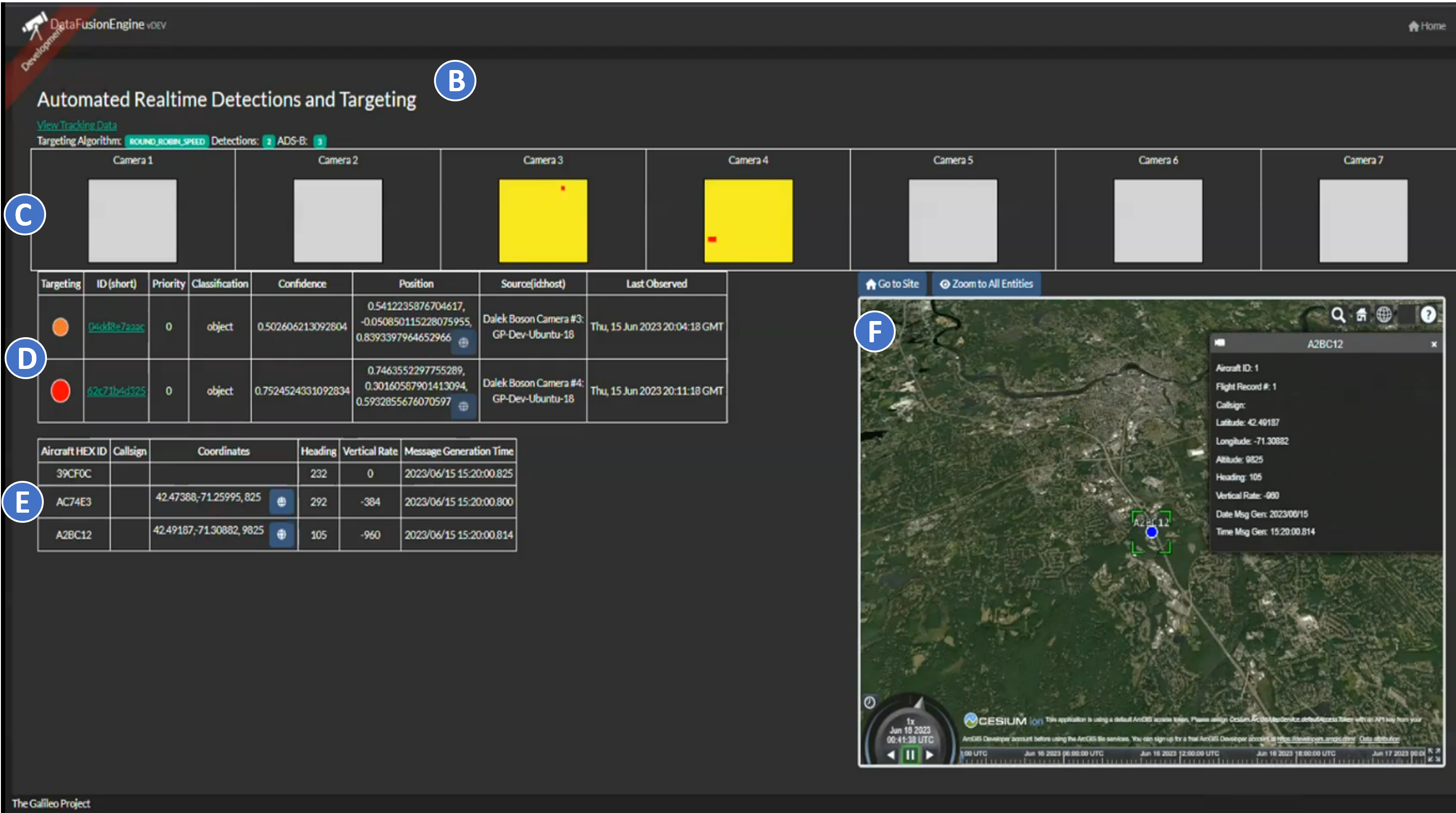}
\end{adjustwidth}
\caption{Data Fusion Engine (DFE) Dashboard screen capture illustrating real-time detection and tracking capabilities. Annotated features include pre-configured dashboards (B), graphical camera layout with detection indicators (C), target-linked ADS-B data (D), detection camera record tables (E), and geospatial visualization of detected entities (F). The dashboard supports role-based access and is built using CesiumJS for 3D geospatial rendering.\label{fig8}}
\end{figure} 

In general, the DFE Dashboard provides real-time situational understanding of the current observable volume of interest, and, in particular, how the DFE data collection system interprets it.  As more sensor types and modalities come online, a dashboard such as this will require additional visualization strategies.  The DFE Dashboard is implemented using widely available Javascript libraries.  The geospatial display is built using CesiumJS \cite{cesiumgs_cesiumjs_2011}.

The DFE publishes detection events for detection to support the ability to monitor events in near real-time.  These capabilities are represented by Quick Look Data Product Generation and Evidence Review Tool, respectively.  Quick Look Data Product is a lightweight summary of raw data for inspection purposes. Current development plans for data product generation go beyond Quick Look Data Products and will utilize the HDF5 format \cite{the_hdf_group_hierarchical_1997} to provide a comprehensive and structured data product.

\begin{table}[H] 
\caption{Structured data product for lightweight summary of raw data for inspection purposes.\label{tab4}}
\begin{tabularx}{\textwidth}{>{\raggedright\arraybackslash}p{4cm} >
{\raggedright\arraybackslash}X}

\toprule
\textbf{Data Category, /Subcategory}	&  \textbf{Description}\\
\midrule
Instrument Science Data &Data originating from sensor(s) within a scientific instrument which may be raw, or processed into calibrated measurands.\\
\hline
Raw Data &Data from any source that has not yet been processed or reduced into a set of  calibrated measurands.\\
\hline
Reduced Data   &Data from any source that has been processed into a set calibrated measurands and to remove errors, duplicate data etc.\\
\hline
\hspace{0.5cm}/Quicklook Data   &A data product or set of products derived from instrument science data that is intended to provide science team members preliminary results from an observational or experimental run.\\
\hline
\hspace{0.5cm}/Non-Quicklook Data &An optional reduced data product for supplementing Quicklook data.\\
\hline
Observatory Site Monitoring Data &Data that is observed remotely (in real-time) that reflect the health and operational status of an observatory's science instruments, systems, components, physical and IT infrastructure, etc.\\
\hline
Raw Health and Status &All data that conveys the health and operational status of an observatory's science instruments, systems, components, physical and IT infrastructure, etc. but has not undergone data reduction nor calibration.\\
\hline
Calibrated Health and Status &All data that conveys the health and operational status of an observatory's science instruments, systems, components, physical and IT infrastructure, etc. that has undergone data reduction and calibration.\\
\hline
Configuration Snapshots &A collection of data at the same epoch that is focused solely on conveying the configuration of a system, component, instrument or even the entire observatory.\\
\hline
Event Messages &Data that conveys the occurrence of a specific event, typically in the form of text or a code associated with that event.\\
\hline
\hspace{0.5cm}/Alerts and Alarms &Typically a message in the from of a GUI/screen indicator often accompanied by audible tones or messages that is intended to highlight to operators an unsafe or potentially unsafe condition in instruments, systems, components, etc. that requires near-term monitoring or action to remedy.\\
\hline
\hspace{0.5cm}/Informational Messages &Typically a message in the from of a GUI/screen indicator often accompanied by audible tones or messages that is intended to highlight to operators actions, conditions, configurations, or events in instruments, systems, components, etc. that have been determined to be of interest to operators.\\
\hline
System Logs &Data that is typically in the form of text messages conveying lower level operational actions, conditions, or events of observatory systems and components.\\
\hline
Detection Event Data &An aggregate of raw and reduced instrument data, quicklook data, and associated ancillary data such as health and status, event messages, alerts/alarms, etc. that collectively defines the detection sufficiently to perform detailed scientific analysis.\\
\hline
Non-Outlier Detection &The occurrence of the observatory automation (or post processing data analysis) determining that an observation event is categorized into a defined category with an associated probability.\\
\hline
Outlier Detection &The occurrence of the observatory automation (or post processing data analysis) determining that an observation event is not categorized into a defined category with an associated probability.\\
\hline
Detection List &A list summarizing the detections that have occurred over a given time period.\\
\bottomrule
\end{tabularx} \\
\end{table}

A description of the content of a Quick Look Data Product is shown in the table above.

The OCICP system is designed to support scientific data collection practices and enable collaboration with the open research community. Dr. Nicola Fox, NASA Associate Administrator for the Science Mission Directorate, emphasized that the investigation of UAPs has been notably lacking a structured and transparent approach like the FAIR framework \cite{donaldson_nasa_2023}, \cite{wilkinson_fair_2016}. The OCICP system adopts FAIR-aligned practices, such as using HDF5 formats and implementing a standardized data storage strategy such as the one shown in the accompanying diagram, to ensure data traceability, accessibility, and scientific utility through the generation of Quick Look Data Products.

More details regarding how the Data Inspection features can be used for system performance analysis are provided in the proceeding Post-Processing subsystem discussions.

%% file: sections/2-material-methods/post-processing-design.tex
\pdfoutput=1
As described above, the Post-Processing subsystem is a hybrid computing environment that resides outside of the Observatory area. The Post-Processing subsystem’s goals are to support the data science workflows, namely the Commissioning and Census Operations, the Science Operations, and the System Effectiveness Monitoring Operations. The design and functional decomposition of the Post-Processing subsystem is shown below in Figure~\ref{fig-postprocessing}. 
\begin{figure}[H]
\begin{adjustwidth}{-\extralength}{0cm}
\centering
\includegraphics[width=15.5cm]{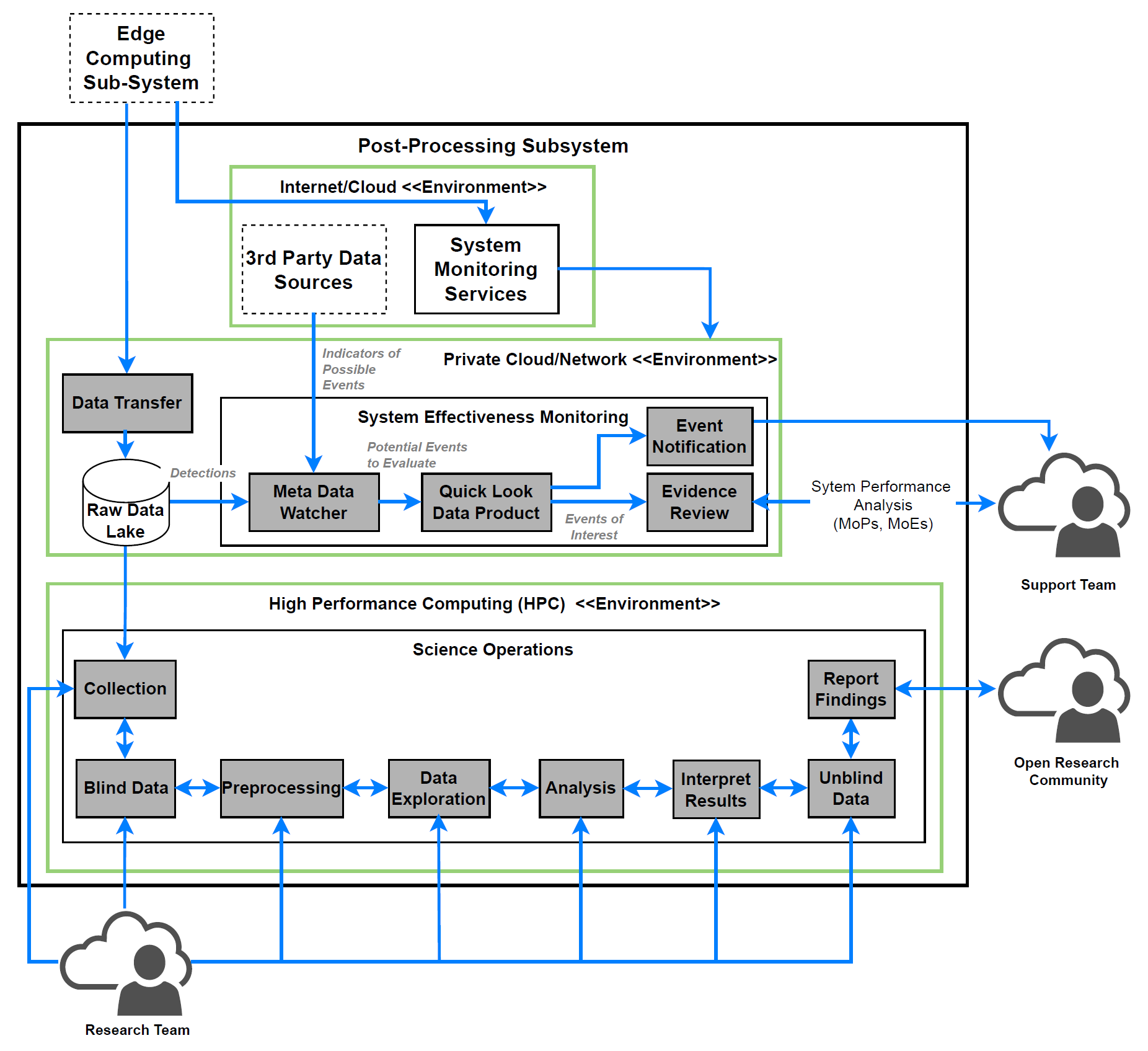}
\end{adjustwidth}
\caption{Post-Processing Subsystem design diagram depicting the functional decomposition of the Post-Processing Subsystem and the primary interactions. \label{fig-postprocessing}}
\end{figure} 
The Post-Processing subsystem is named as such because this subsystem supports all processing and workflows that occur after the initial data collection and data reduction of the Edge Computing subsystem.  The Post-Processing subsystem is defined as a hybrid computing environment because it is composed of a combination of Internet/cloud, private network, and HPC (High Performance Computing) environments. The use of these environments, and the locations of specific functionality, is indicated in the design diagram with the stereotype label ``<<Environment>>" and shown outlined in green boxes.  The following subsections will describe each of these environments in more detail.

\subsubsection{Internet/Cloud Environment}
The internet provides multiple services to support team collaboration, software and configuration version control, programmatic services, and system monitoring services.  For the purpose of science operations, the Internet / cloud environment provides the opportunity to leverage querying of regional sensor networks such as ADS-B transponder data, atmospheric data, ultrasound, and magnetometers to distinguish events with effects that are purely local from those whose effects are regional, as described in Watters et al. \cite{watters_scientific_2023}.

\subsubsection{Private Network Environment}
Several key services reside within the private network environment. Utilizing a private network provides the opportunity to stand-up and self-host services in a way that only requires one-time purchases (e.g., purchasing a storage device), as well as allows for the utilization of preexisting research organization resources. This environment cost model is a good fit for a donor-funded research group such as this, compared to committing to ongoing costs associated with third-party services.  For the most part, this environment is utilized to host persistent services and data stores that are usable across all observatories.

The first activity that must be performed within the Post-Processing Computing subsystem is to acquire the data from the Edge Computing subsystem. The Data Transfer software-based service provides the functionality of transferring data from the Edge Computing subsystem into the Post-Processing subsystem. This service represents multiple data transfer utilities, such as partial transfers based on sensor type and date filtering, complete transfers, and data comparison tools. This service may be executed from multiple environments, but when executed in an automated fashion (e.g., scheduled), it resides within the Private Network Environment. Due to large data volumes and communications bandwidth constraints for remote observatory sites, we expect to utilize manual data pickup and transfer processes. However, once data are retrieved and available on a personal computer, the Data Transfer software service will assist in the transfer of data to the private cloud environment.

The Raw Data Lake is the primary data storage used for maintaining the collected sensor data and related meta-data.  The reason why this data storage is defined as a Data Lake is that this data store contains various unprocessed raw data types, compared to a Data Mart, which is filtered, processed, and structured for specific purposes \cite{inmon_data_2016}. Per the goal of the Edge Computing subsystem, Management of Data Provenance, the Edge Computing subsystem provides the requisite information required to fully track and assess the provenance and traceability of sensor data, along with relevant calibration data. Furthermore, an important attribute of this data lake is that these data are immutable. The Raw Data Lake provides data for two essential operations: System Effectiveness Monitoring and Science Operations. 

The System Effectiveness Monitoring operation supports the ability for the Support Team to be notified of detection events and be able to assess the performance and effectiveness of the system. In context of this function, the assessment is focused on the Edge Computing subsystem and quality of data within the Raw Data Lake.  As depicted in the diagram, the output of the System Effectiveness monitoring process influences the science operations process.  

Several core capabilities make up the System Effectiveness Monitoring: Meta Data Watcher, Quick Look Data Product, Event Notification, and Evidence Review. The Meta Data Watcher is responsible for monitoring data (for example, log files) for Edge Computing subsystem detections, as well as external data sources that indicate potential opportunities for aerial observations with an observatory’s detection volume. The Evidence Review function provides the means to efficiently analyze and review data.  This function also attempts to simplify the use of multimodal review tools, such as synchronized video playback, audio playback, and spectral visualization. When the Support Team receives an event notification, the notification will indicate where the prepared data package resides, and at this point the Support Team can utilize the Evidence Review tool, as well as other ad-hoc means, to analyze the data. The outcome of this workflow is to determine the performance and effectiveness of the system and not to influence the execution of the Science Operations.

These events represent the opportunity to direct the System Effectiveness Monitoring workflow and Support Team to particular data of interest.  Data of interest for the purpose of monitoring system effectiveness represent prosaic events and not indicators of UAP. These data and the notification event represent an opportunity to examine the performance and behavior of the observatory.  The goal of evaluating these events includes opportunities to inspect possible system failures, anomalies in mistreatment performance, and more.

\subsubsection{High Performance Computing Environment (HPC)}
The High-Performance Computing Environment, or HPC, is an environment designed to efficiently perform complex processing problems that require large amounts of data. The HPC leverages a parallel processing architecture, is highly scalable, and is optimized for high performance by leveraging clusters of high-powered processors. 

The HPC infrastructure is required by the Science Operations data science activities.  The activities of Collection, Pre-processing, Data Exploration, Analysis, Interpret Results, and Report Findings represent the general activities performed within data science workflows. Unique to our approach to UAP investigation is the application of data blinding, described in Section~\ref{sec:data_blinding}. The Research Team performs all of the Science Operations activities.  Although the observatory site locations are not made available to the general public, the results of this workflow are shared with the Open Research Community via the Research Team.  Examples of results sharing include published peer-reviewed science papers and the sharing of research data.

%% file: sections/2-material-methods/post-processing-implementation.tex
\pdfoutput=1
\subsection{Post-Processing Subsystem Implementation}
This section provides a more detailed view of the Post-Processing subsystem, with a focus on its implementation and component interactions to illustrate the system's dynamic behavior. Some implementation features described in this section may not be in use or even operational due to the fluid nature of scientific project requirements, execution, and priorities; however, all implementation details described in this section represent capabilities that have been developed and have been used at some time, unless otherwise specified as 'planned'. 
\begin{figure}[H]
\begin{adjustwidth}{-\extralength}{0cm}
\centering
\includegraphics[width=15.5cm]{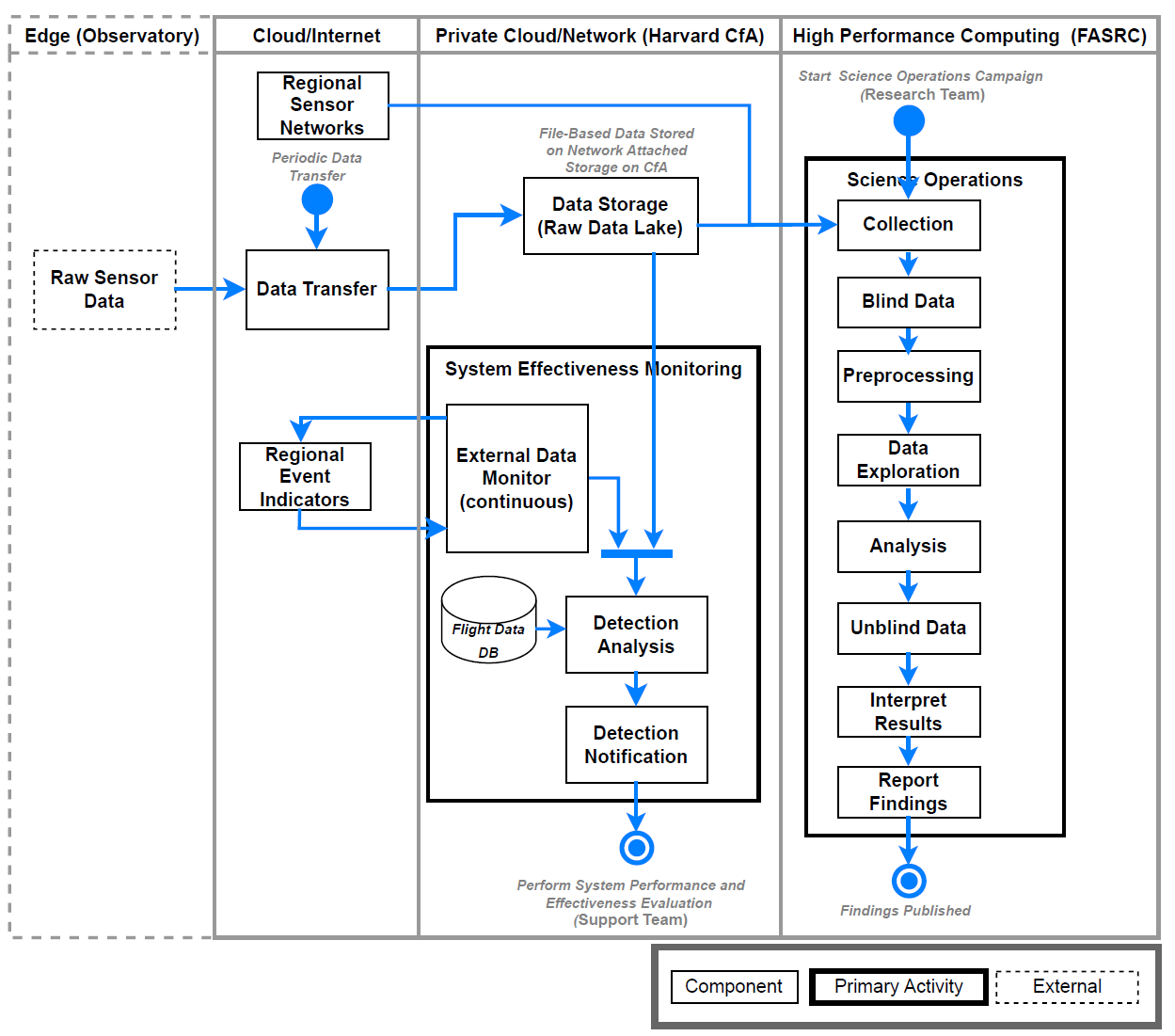}
\end{adjustwidth}
\caption{Post-Processing Subsystem Component interaction diagram depicting the implemented components and primary interactions, organized by capability categories.
\label{fig9}}
\end{figure} 
The above interaction diagram depicts the Postprocessing subsystem services as primary components (container boxes with bold titles), components (boxes), and component interactions (arrows), all of which are organized within computing environments (columns).  

\subsubsection{Private Network}
As the left side of the diagram, sensor data is retrieved from the Edge Computing subsystem by the data transfer component. Data transfers are performed periodically according to a schedule that provides data to the Post-Processing subsystem when needed, as well as reduce risk of data loss within the Edge Computing subsystem.  Because observatories may be remote, the Data Transfer service may consist of multiple implementations.  The current implementation is primarily a manual process supported by custom data transfer script tools and the physical transfer of removable hard drives.
The Raw Data Lake is file-based and hosted in Harvard's Faculty of Arts and Sciences Division of Science Research Computing high performance computing environment, also known as FASRC {\cite{noauthor_fas_nodate}} 

\subsubsection{Science Operations: Data Collection}
The Data Collection activity is focused on gathering relevant data for analysis. For the Post-Processing Subsystem, data can come from either the Raw Data Lake or Regional Sensor Networks. This stage ensures that the collected data aligns with the research question or the objectives of the project. Considerations for this activity include how to access, store, and manage the collected data, as well as the data format and sampling methods. 
With the requirement for the use of a multimodal and multispectral approach, employing various instruments and sensors to collect data from diverse sources \cite{watters_scientific_2023}, the Data Collection activity will be required to collect a diverse set of heterogeneous data as more sensors and sensor systems complete the commissioning process and begin to produce science-quality data. Watters et al. \cite{watters_scientific_2023} also highlights the importance of careful site selection to optimize data collection, considering factors such as geographic location, atmospheric conditions, and potential for UAP sightings. Refer to Watters et al. for a detailed discussion of the relevant instrumentation and data collection strategies \cite{watters_scientific_2023}. 

\subsubsection{Science Operations: Blind Data}
\label{sec:data_blinding}
Once all instruments in the observatory have been through a commissioning phase, and begin collecting science-quality data, we plan to enable a data blinding scheme \cite{maccoun2015blind, klein2005blind}. For a given data collection time period, the real values of certain key measurements will be obfuscated with a lossless encryption, which means that the data preprocessing, exploration, and analysis steps will be performed blindly. For example, if one of the measurements from the passive radar instrument is the speed for each detected object, the distribution of speed measurements can be normalized; the mean and standard deviation of the original distributions, key numbers to invert this transformation, would be stored safely away until the decision to unblind is taken. Blind analysis helps to avoid bias from the experimenter's own preconceptions, which could lead to confirming prior beliefs. All analysis decisions must be made and all code troubleshooting completed before the data bucket for that time period is allowed to be ``unblinded", or decrypted, and the results can then be interpreted.

\subsubsection{Science Operations: Data Preprocessing}
The Data Preprocessing activity is focused on preparing raw data for analysis by cleaning and transforming them. This activity addressed the need to deal with data that contain inconsistencies, missing values, or errors. Preprocessing involves handling these issues by removing duplicates, normalizing or scaling data, and encoding categorical variables. This step ensures that the data are clean, accurate, and formatted to be usable in the analysis stage, improving the quality of the results.

The Data Exploration activity focuses on understanding latent data patterns and relationships within the data to guide further analysis. With the application of techniques such as descriptive statistics and data visualization, data exploration provides insights into the structure, distribution, and possible trends within the data. Analysts may also identify potential correlations, anomalies, or outliers within the data set. This stage helps refine the hypotheses and adjust the planned analysis approach based on preliminary findings.

\subsubsection{Science Operations: Data Analysis}
The Data Analysis activity is focused on the application of statistical and machine learning methods to extract information from the data. In this phase, advanced analytical techniques are applied to investigate specific hypotheses or answer research questions. 

\subsubsection{Science Operations: Interpret Results}
The Interpret Results activity is focused on drawing meaningful conclusions from the Data Analysis results. After Data Analysis, the results are interpreted within the context of the scientific study of UAP. Determining the implications, limitations, and reliability of the findings is part of this activity.

As the OCICP system continues to mature and as more instruments are utilized, the OCICP system will provide additional support for outlier detection and the Science Operations phase at large.  

\subsubsection{Science Operations: Report Findings}
The Report Findings activity focuses on communicating insights and findings to the Open Research Community, as well as supporting the commitment to data transparency and the publication of findings in peer-reviewed journals.

%% file: sections/3-results/results.tex
\pdfoutput=1
The Observatory Class Integrated Computing Platform (OCICP) has been successfully deployed at three observatory sites and is actively supporting long-term, multimodal data collection for the scientific study of Unidentified Anomalous Phenomena (UAP). These deployments demonstrate that OCICP is a viable and scalable architecture for sustained, transparent, and scientifically rigorous monitoring operations.

\subsection{Implementation Status}
OCICP is being deployed at multiple observatory sites and is in different stages of setup and operational status, each observatory site having a distinct commissioning phase that must be completed before Science Operations can begin. While the system architecture is designed to support capabilities like real-time object detection, tracking, and data fusion through its Edge Computing subsystem, and features for system effectiveness monitoring via data inspection, some implemented features may not be in use currently, or even operational. For example, real-time classification of objects is not currently implemented, and autonomous sensor response and expanded real-time decision-making capabilities are planned for future work. The current focus is on conducting the aerial survey and post-processing activities, utilizing the collected data within the High-Performance Computing environment for Science Operations, including data collection, pre-processing, analysis, and interpretation. 

\subsection{Example data collected by OCICP}

\begin{figure}[H]
\centering
\centering
\includegraphics[width=10 cm]{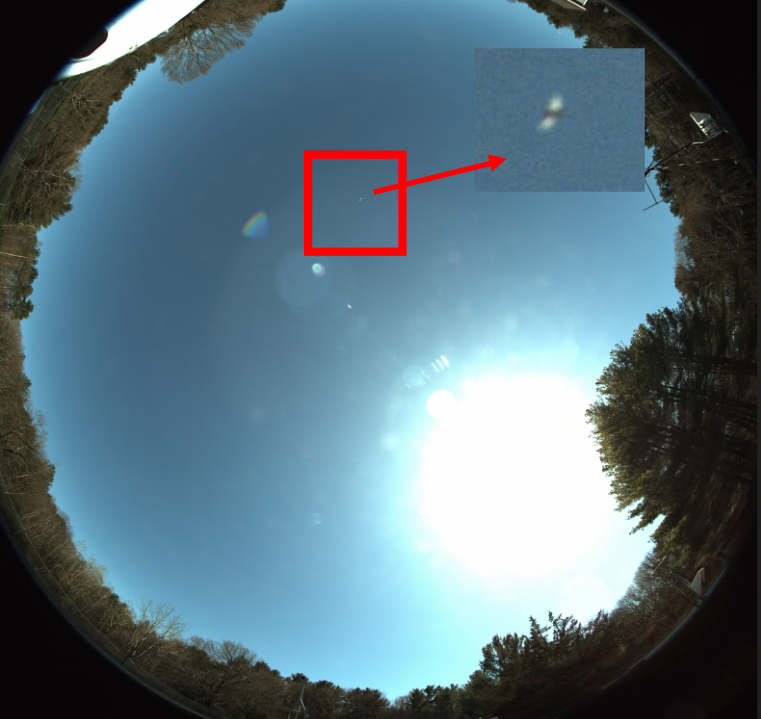}
\caption{This image was captured using the Alcor all-sky imaging system, equipped with a high-resolution sensor producing 36-megapixel (6000 × 6000 pixel) hemispherical sky views.\label{fig:alcor}}
\end{figure}

The Alcor is an all-sky imaging system which captures high-resolution hemispherical images of the sky using a 36-megapixel sensor, producing individual files of approximately 5 MB in size. These images are collected at a rate of 0.5 to 1 frame per second, with dynamic exposure settings optimized for diurnal and nocturnal conditions. The resulting dataset provides near-continuous visual coverage of the full celestial dome, allowing for the detection and tracking of aerial objects, cloud movement, and atmospheric phenomena. Figure~\ref{fig:alcor} shows an example of Alcor image.

The Dalek infrared camera system is a multi-sensor array composed of eight infrared cameras. Each camera operates at a resolution of 640 × 512 pixels and records at 10 frames per second, producing synchronized video streams for object detection and tracking. Collectively, the Dalek array generates approximately 90 GB of video data per day. The dataset includes detailed metadata for each video, including timestamps and calibrated camera pose information. All cameras were intrinsically calibrated to account for lens distortion and internal optical parameters, and extrinsically calibrated to define their spatial orientation and relative positioning within the array. 

\begin{figure}[H]
\centering
\includegraphics[width=\linewidth]{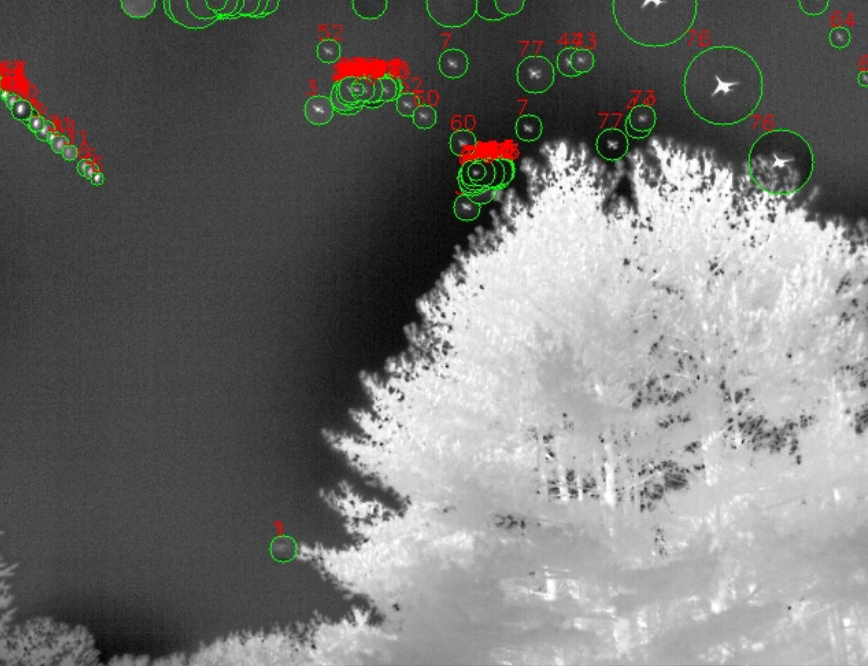}
\caption{This image shows infrared detections made by the Dalek system for a 5-minute video, using frame-by-frame YOLOv5 object recognition. The detected objects are enclosed in a green bounding box and annotated with a red identifier number. They are overlaid on top of a background frame, drawn from a moment in the same video without active detections, which provides environmental context for the detection.\label{fig2}}
\end{figure} 

The raw videos are processed for object detection (based on YOLOv5), localization, and trajectory reconstruction. Figure~\ref{fig2} contains an example summary of object detections and trajectories in one video. Post-processing includes treeline masking to eliminate false positives and spline smoothing to refine trajectories. 

\begin{figure}[H]
\centering
\includegraphics[width=10.5cm]{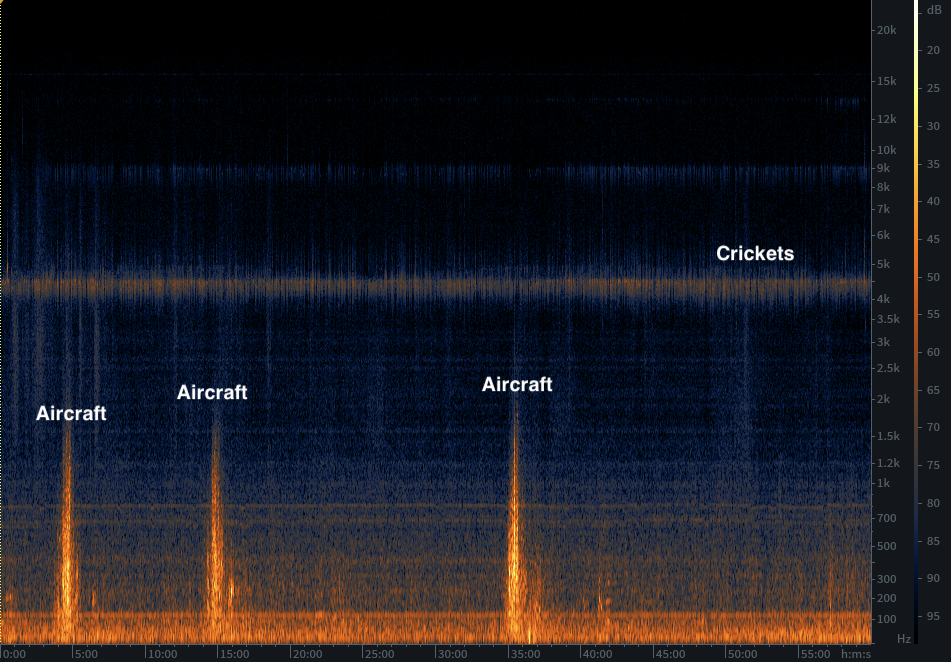}
\caption{This spectrogram reveals the frequencies in a two-minute uncompressed .wav file recorded by the AMOS system. The x-axis represents time (hh:mm:ss), the y-axis shows frequency in Hz, and color intensity represents signal strength in decibels (dB).\label{fig3}}
\end{figure}

The acoustic system captures high-fidelity environmental audio using a GRAS 41AC3 sensor paired with a HiFiBerry analog-to-digital converter. The resulting dataset consists of raw, uncompressed .wav files accompanied by spectrograms and spectral energy profiles. Each file represents a continuous sixty-minute audio segment, enabling fine-grained temporal analysis of acoustic signatures.
Figure~\ref{fig3} is an example of an acoustic spectrogram recorded at one of the observatories.  Post-processing is uses spectral denoising algorithms and normalization routines (e.g., Izotope RX) to enhance signal clarity while retaining authentic sound structure.
The acoustic dataset is an important component in the multi-sensor data fusion, offering an independent channel for cross-referencing objects detected in optical sensors or other modalities.

A geomagnetic variometer collects continuous measurements of the ambient magnetic field using a high-sensitivity, three-axis fluxgate magnetometer (Bartington Mag-13MS100). The data are recorded at a sampling rate of 1612.9 Hz, enabling the detection of magnetic field variations across a broad frequency spectrum up to 800 Hz. The system records vector field components (X, Y, Z) with nanotesla (nT) sensitivity and is equipped with temperature calibration to account for thermal drift.

\begin{figure}[H]
\centering
\includegraphics[width=10.0cm]{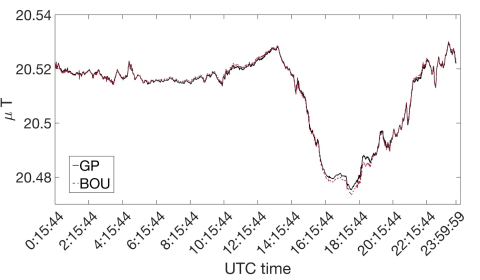}
\caption{This plot presents a side-by-side comparison of magnetic field strength variations recorded independently by a Galileo Project (GP) magnetometer and the USGS Boulder observatory (BOU) over a 24-hour period. Both datasets were sampled at 1-second intervals and aligned through baseline adjustment to ensure comparability.\label{fig4}}
\end{figure}

This dataset captures both naturally occurring geomagnetic activity—such as diurnal variations and geomagnetic storms—as well as potential transient anomalies of interest in UAP investigations. All data are stored in one-hour segments and are downsampled to 1-second intervals for cross-comparison with regional observatory data (see Figure~\ref{fig4} for example). Calibration was performed in collaboration with the USGS magnetic observatory in Boulder, Colorado, ensuring that the GP system meets INTERMAGNET-quality standards. This alignment provides confidence in anomaly detection by validating that the GP system’s magnetic measurements are consistent with those from established geophysical observatories.

ADS-B (Automatic Dependent Surveillance–Broadcast) is a near-real-time aircraft positioning system in which aircraft regularly broadcast their identity, GPS-derived position (latitude, longitude, altitude), velocity, and heading. Each message includes a timestamp and a unique ICAO address for the aircraft, allowing for unambiguous identification. ADS-B is vital for correlating known aerial traffic with object detections.
During a five-month data collection period, the ADS-B receiver registered approximately 27,500 aircraft passing within a 10 km radius of the observatory. On average, about 200 ADS-B-equipped aircraft were detected daily within this range. 

\subsection{System Performance and Data Yield}
OCICP currently generates approximately 350 GB of multimodal data per day, a majority of which originate from the Dalek infrared camera array and the Alcor all-sky camera. Each Dalek camera contributes to a combined 90 GB daily data yield, while the Alcor camera contributes approximately 215 GB. Acoustic sensors produce around 8 GB per day, and additional meteorological and magnetometric data contribute a smaller data volume as structured time series.

\subsection{Operational Challenges and Mitigation Strategies}
Field deployments revealed several challenges, such as adverse weather conditions, limited network connectivity, and dynamic obstructions (e.g., vegetation growth). These factors impact image clarity, instrument/sensor up-time, and event reliability. To mitigate these, OCICP employs treeline masking, automated sunshield deployment for infrared sensors, and local data persistence mechanisms. The modular design of the architecture and the event-driven data routing ensure the resilience and adaptability of the system to changing operational environments.

\subsection{System Performance scaling}
The OCICP system architecture uses two distinct subsystems to optimize performance and scaling.  The first is the Edge Computing Sub-System and the second is the Post-Processing Sub-System. The Post-Processing Sub-System utilizes cloud and high-performance computing environments which inherently support scaling by leveraging distributed architectures and dynamic resource orchestration to provision compute, storage, and networking on demand, enabling parallel execution and rapid elasticity to match workload demands.  This leaves us with addressing scaling and performance for the Edge Computing Sub-System.

The primary strategy applied for the Edge Computing Sub-System is to leverage an Event-Driven Architecture which enables the decoupling of components via asynchronous events and messaging, allowing services to scale independently and process high-throughput workloads in parallel for optimal performance. Building on this messaging backbone, the sub-system leverages microservices to perform distinct functions. A microservice is a small, independently deployable component that encapsulates a single functional capability, and this modularity allows each service to be scaled and optimized for performance without impacting the rest of the system. An example of this is the application of data collectors as small message-based microservices that are capable of collecting raw sensor data and making those data available to the system in a modular and scalable manner via the message bus.

Some data acquisition schedules are dynamically adjusted to mitigate environmental challenges, for example exposure to sunlight for infrared sensors. Operational integrity and uptime are continuously monitored through a site monitoring platform, and all collected data are initially stored on site via removable solid-state drives before being transmitted to a centralized repository for further analysis. Although the OCICP architecture has been designed to support capabilities such as PTZ tracking and radar integration, these specific features have not yet been field-tested or operationally validated at the time of writing.

\subsection{Sensor Integration and Cross-Modality Observations}
OCICP's architecture enables concurrent operation of heterogeneous sensors with seamless data provenance and calibration traceability. Each sensor—Dalek (IR optical), Alcor (visible spectrum all-sky), AMOS (acoutstic monitoring), magnetometer, ADS-B receiver, and audio (acoustic and ultrasound)—provides independent but integrable observational streams. Through the Edge Computing subsystem, OCICP applies event-driven processing and multi-level data fusion (JDL model), enabling real-time object detection, trajectory reconstruction, and multimodal entity resolution. This integration allows for robust cross-verification of aerial detections with environmental and contextual metadata, significantly improving the confidence level in event classification.

\subsection{Scientific Utility and Transparency}
An important motivation of this study is that existing military grade systems are not suitable for open scientific inquiry due to classification constraints. OCICP addresses this gap directly by offering a fully transparent open-data alternative with standardized calibration, immutable provenance, and reproducible workflows. All data products, including raw, quick-look and analysis-ready formats (e.g. HDF5), are traceable and FAIR-compliant, supporting collaboration across the broader research community. The availability of open-source environmental datasets (e.g., OpenMeteo) is able to further enhance the utility of the OCICP system by enriching the event context and enabling third-party validation.

\subsection{Future Research Directions}
The current OCICP deployment establishes a foundation for performing aerial surveys and supporting science operations. Future development will prioritize real-time classification pipelines using ensemble machine learning strategies within the Data Fusion Engine (DFE), expanded use of controllable sensors such as PTZ cameras, and integration of additional modalities (e.g., passive radar, optical spectroscopy).

%% file: template.bbl
\begin{thebibliography}{999}

\bibitem[Watters et~al.(2023)Watters, Loeb, Laukien, Cloete, Delacroix, Dobroshinsky, Horvath, Kelderman, Little, Masson, Mead, Randall, Schultz, Szenher, Vervelidou, White, Ahlström, Cleland, Dockal, Donahue, Elowitz, Ezell, Gersznowicz, Gold, Hercz, Keto, Knuth, Lux, Melnick, Moro-Martín, Martin-Torres, Ribes, Sail, Teodorani, Tedesco, Tedesco, Tu, and Zorzano]{watters_scientific_2023}
Watters, W.A.; Loeb, A.; Laukien, F.; Cloete, R.; Delacroix, A.; Dobroshinsky, S.; Horvath, B.; Kelderman, E.; Little, S.; Masson, E.;  et~al.
\newblock The {Scientific} {Investigation} of {Unidentified} {Aerial} {Phenomena} ({UAP}) {Using} {Multimodal} {Ground}-{Based} {Observatories}.
\newblock {\em Journal of Astronomical Instrumentation} {\bf 2023}, p. 2340006.
\newblock Publisher: World Scientific Publishing Co., {\url{https://doi.org/10.1142/S2251171723400068}}.

\bibitem[Loeb and Laukien(2022)]{loeb_overview_2022}
Loeb, A.; Laukien, F.H.
\newblock Overview of the {Galileo} {Project}.
\newblock {\em Journal of Astronomical Instrumentation} {\bf 2022}, p. 2340003.
\newblock Publisher: World Scientific Publishing Co., {\url{https://doi.org/10.1142/S2251171723400032}}.

\bibitem[noa(2024)]{noauthor_seti_2024}
{SETI} {Institute},  2024.

\bibitem[Cloete et~al.(2023)Cloete, Bridgham, Dobroshinsky, Ezell, Fedorenko, Laukien, Little, Loeb, Masson, Szenher, and Watters]{cloete_integrated_2023}
Cloete, R.; Bridgham, P.; Dobroshinsky, S.; Ezell, C.; Fedorenko, A.; Laukien, F.; Little, S.; Loeb, A.; Masson, E.; Szenher, M.;  et~al.
\newblock Integrated {Computing} {Platform} for {Detection} and {Tracking} of {Unidentified} {Aerial} {Phenomena} ({UAP}).
\newblock {\em Journal of Astronomical Instrumentation} {\bf 2023}.
\newblock Publisher: World Scientific Publishing Co., {\url{https://doi.org/10.1142/S2251171723400081}}.

\bibitem[noa(2024)]{noauthor_omg_2024}
{OMG} {Systems} {Modeling} {Language},  2024.

\bibitem[NASA(2023)]{nasa_nasa_2023}
NASA.
\newblock {NASA} {Unidentified} {Anomalous} {Phenomena} {Independent} {Study} {Team} {Report},  2023.

\bibitem[noa(2023)]{noauthor_dod_2023}
{DOD} {Official} {Testifies} {Before} {Senate} {Subcommittee},  2023.

\bibitem[Vergun(2023)]{vergun_dod_2023}
Vergun, D.
\newblock {DOD} {Working} to {Better} {Understand}, {Resolve} {Anomalous} {Phenomena},  2023.

\bibitem[Donaldson(2023)]{donaldson_nasa_2023}
Donaldson, A.
\newblock {NASA} {Provides} {Coverage} of {Unidentified} {Anomalous} {Phenomena} {Meeting},  2023.

\bibitem[Kayal et~al.(2023)Kayal, Greiner, Kaiser, and Riegler]{kayal_erforschung_2023}
Kayal, H.; Greiner, T.; Kaiser, T.; Riegler, C.
\newblock Erforschung des {Unidentified} {Aerial} {Phenomena} an der {JMU} {Würzburg},  2023.

\bibitem[Szydagis et~al.(2024)Szydagis, Knuth, Kugielsky, and Levy]{szydagis_initial_2024}
Szydagis, M.; Knuth, K.H.; Kugielsky, B.W.; Levy, C.
\newblock Initial {Results} {From} the {First} {Field} {Expedition} of {UAPx} to {Study} {Unidentified} {Anomalous} {Phenomena},  2024.
\newblock arXiv:2312.00558.

\bibitem[Domine et~al.(2024)Domine, Biswas, Cloete, Delacroix, Fedorenko, Jacaruso, Kelderman, Keto, Little, Loeb, Masson, Schultz, Szenher, Watters, and White]{domine_commissioning_2024}
Domine, L.; Biswas, A.; Cloete, R.; Delacroix, A.; Fedorenko, A.; Jacaruso, L.; Kelderman, E.; Keto, E.; Little, S.; Loeb, A.;  et~al.
\newblock Commissioning {Galileo} {Project}'s {All}-{Sky} {Infrared} {Camera} {Array} for {Aerial} {Detections}.
\newblock {\em Sensors} {\bf 2024}.

\bibitem[Luckham(2002)]{luckham_power_2002}
Luckham, D.
\newblock {\em The power of events}; Vol. 204, Addison-Wesley Reading,  2002.

\bibitem[Steinberg et~al.(1999)Steinberg, Bowman, and White]{steinberg_revisions_1999}
Steinberg, A.N.; Bowman, C.L.; White, F.E.
\newblock Revisions to the {JDL} data fusion model.
\newblock In Proceedings of the Sensor {Fusion}: {Architectures}, {Algorithms}, and {Applications} {III}. SPIE,  March 1999, Vol. 3719, pp. 430--441.
\newblock {\url{https://doi.org/10.1117/12.341367}}.

\bibitem[II et~al.(2017)II, Hall, and Llinas]{ii_handbook_2017}
II, M.L.; Hall, D.; Llinas, J.
\newblock {\em Handbook of {Multisensor} {Data} {Fusion}: {Theory} and {Practice}, {Second} {Edition}}; CRC Press,  2017.

\bibitem[{ZeroMQ}(2023)]{zeromq_zeromq_2023}
{ZeroMQ}.
\newblock {ZeroMQ},  2023.
\newblock Programmers: \_:n12442.

\bibitem[noa(2024)]{noauthor_openjdk_2024}
{OpenJDK},  2024.

\bibitem[Randall et~al.(2022)Randall, Delacroix, Ezell, Kelderman, Little, Loeb, Masson, Watters, Cloete, and White]{randall_skywatch_2022}
Randall, M.; Delacroix, A.; Ezell, C.; Kelderman, E.; Little, S.; Loeb, A.; Masson, E.; Watters, W.A.; Cloete, R.; White, A.
\newblock {SkyWatch}: {A} {Passive} {Multistatic} {Radar} {Network} for the {Measurement} of {Object} {Position} and {Velocity}.
\newblock {\em Journal of Astronomical Instrumentation} {\bf 2022}, p. 2340004.
\newblock Publisher: World Scientific Publishing Co., {\url{https://doi.org/10.1142/S2251171723400044}}.

\bibitem[{FAA}(2023)]{faa_automatic_2023}
{FAA}.
\newblock Automatic {Dependent} {Surveillance}-{Broadcast} ({ADS}-{B}) {\textbar} {Federal} {Aviation} {Administration},  2023.

\bibitem[{Raspberry Pi}(2024)]{raspberry_pi_raspberry_2024}
{Raspberry Pi}.
\newblock Raspberry {Pi} 4 {Model} {B},  2024.

\bibitem[noa(2024)]{noauthor_flightaware_2024}
{FlightAware} {Pro} {Stick}® {Plus} {\textbar} {ADS}-{B} {SDR} {USB} dongle,  2024.

\bibitem[Szenher et~al.(2022)Szenher, Delacroix, Keto, Little, Randall, Watters, Masson, and Cloete]{szenher_hardware_2022}
Szenher, M.; Delacroix, A.; Keto, E.; Little, S.; Randall, M.; Watters, W.A.; Masson, E.; Cloete, R.
\newblock A {Hardware} and {Software} {Platform} for {Aerial} {Object} {Localization}.
\newblock {\em Journal of Astronomical Instrumentation} {\bf 2022}, p. 2340002.
\newblock Publisher: World Scientific Publishing Co., {\url{https://doi.org/10.1142/S2251171723400020}}.

\bibitem[noa()]{noauthor_skyalert_nodate}
{SkyAlert}.

\bibitem[Jocher(2020)]{jocher_yolov5_2020}
Jocher, G.
\newblock {YOLOv5},  2020.
\newblock Publication Title: GitHub repository.

\bibitem[noa(2024)]{noauthor_gstreamer_2024}
{GStreamer}: open source multimedia framework,  2024.

\bibitem[{Python Software Foundation}(2008)]{python_software_foundation_python_2008}
{Python Software Foundation}.
\newblock Python 3.0,  2008.

\bibitem[Pivotal~Software(2014)]{pivotal_software_spring_2014}
Pivotal~Software, I.
\newblock Spring {Boot},  2014.

\bibitem[Hohpe and Woolf(2003)]{hohpe_enterprise_2003}
Hohpe, G.; Woolf, B.
\newblock {\em Enterprise {Integration} {Patterns}: {Designing}, {Building}, and {Deploying} {Messaging} {Solutions}}; Addison-Wesley Professional,  2003.

\bibitem[Gamma et~al.(1994)Gamma, Helm, Johnson, and Vlissides]{gamma_design_1994}
Gamma, E.; Helm, R.; Johnson, R.; Vlissides, J.
\newblock {\em Design {Patterns}: {Elements} of {Reusable} {Object}-{Oriented} {Software}}; Addison-Wesley Professional: Boston, MA, USA,  1994.

\bibitem[Alur et~al.(2003)Alur, Crupi, and Malks]{alur_core_2003}
Alur, D.; Crupi, J.; Malks, D.
\newblock {\em Core {J2EE} {Patterns}: {Best} {Practices} and {Design} {Strategies}}, 2nd edition ed.; Prentice Hall PTR,  2003.

\bibitem[{CesiumGS}(2011)]{cesiumgs_cesiumjs_2011}
{CesiumGS}.
\newblock {CesiumJS},  2011.

\bibitem[{The HDF Group}(1997)]{the_hdf_group_hierarchical_1997}
{The HDF Group}.
\newblock Hierarchical {Data} {Format}, version 5,  1997.

\bibitem[Wilkinson et~al.(2016)Wilkinson, Dumontier, Aalbersberg, Appleton, Axton, Baak, Blomberg, Boiten, Santos, Bourne, and {others}]{wilkinson_fair_2016}
Wilkinson, M.D.; Dumontier, M.; Aalbersberg, I.J.; Appleton, G.; Axton, M.; Baak, A.; Blomberg, N.; Boiten, J.W.; Santos, L.B.d.S.; Bourne, P.E.;  et~al.
\newblock The {FAIR} {Guiding} {Principles} for scientific data management and stewardship.
\newblock {\em Scientific Data} {\bf 2016}, {\em 3},~160018.
\newblock Number: 1 Publisher: Nature Publishing Group, {\url{https://doi.org/10.1038/sdata.2016.18}}.

\bibitem[Inmon(2016)]{inmon_data_2016}
Inmon, B.
\newblock {\em Data {Lake} {Architecture}: {Designing} the {Data} {Lake} and avoiding the garbage dump}; Technics Publications, LLC,  2016.

\bibitem[noa()]{noauthor_fas_nodate}
{FAS} {Research} {Computing}.

\bibitem[MacCoun and Perlmutter(2015)]{maccoun2015blind}
MacCoun, R.; Perlmutter, S.
\newblock Blind analysis: Hide results to seek the truth.
\newblock {\em Nature} {\bf 2015}, {\em 526},~187--189.

\bibitem[Klein and Roodman(2005)]{klein2005blind}
Klein, J.R.; Roodman, A.
\newblock Blind analysis in nuclear and particle physics.
\newblock {\em Annu. Rev. Nucl. Part. Sci.} {\bf 2005}, {\em 55},~141--163.

\end{thebibliography}
